\documentclass[twocolumn]{fairmeta}
\usepackage{amssymb}
\usepackage{multirow} 
\usepackage{multirow} 
\usepackage{algorithm}
\usepackage{algpseudocode}
\usepackage{listings}
\usepackage{graphicx}
\usepackage{subcaption}

\title{Bending the Scaling Law Curve in Large-Scale Recommendation Systems}

\author[1,*]{Qin~Ding}
\author[1]{Kevin~Course}
\author[1]{Linjian~Ma}
\author[1]{Jianhui~Sun}
\author[1]{Ruochen~Liu}
\author[1]{Zhao~Zhu}
\author[1]{Chunxing~Yin}
\author[1]{Wei~Li}
\author[1]{Dai~Li}
\author[1]{Yu~Shi}
\author[1]{Xuan~Cao}
\author[1]{Ze~Yang}
\author[1]{Han~Li}
\author[1]{Xing~Liu}
\author[1]{Bi~Xue}
\author[1]{Hongwei~Li}
\author[1]{Rui~Jian}
\author[1]{Daisy~Shi~He}
\author[1]{Jing~Qian}
\author[1]{Matt~Ma}
\author[1]{Qunshu~Zhang}
\author[1,*]{Rui~Li}

\affiliation[1]{Meta Recommendation Systems}
\abstract{
Learning from user interaction history through sequential models has become a cornerstone of large-scale recommender systems. 
Recent advances in large language models have revealed promising scaling laws, sparking a surge of research into long-sequence modeling and deeper architectures for recommendation tasks. 
However, many recent approaches rely heavily on cross-attention mechanisms to address the quadratic computational bottleneck in sequential modeling, which can limit the representational power gained from self-attention.
We present ULTRA-HSTU, a novel sequential recommendation model developed through end-to-end model and system co-design. By innovating in the design of input sequences, sparse attention mechanisms, and model topology, ULTRA-HSTU achieves
substantial improvements in both model quality and efficiency.
Comprehensive benchmarking demonstrates that ULTRA-HSTU achieves remarkable scaling efficiency gains—over $5\times$ faster training scaling and 21$\times$ faster inference scaling compared to conventional models—while delivering superior recommendation quality. 
Our solution is fully deployed at scale, serving billions of users daily and driving significant $4\%$ to $8\%$ consumption and engagement improvements in real-world production environments.
}
\correspondence{\email{qding@meta.com}, \email{ruili@meta.com}}
\date{\today}

\begin{document}

\maketitle

\section{Introduction}

Recently, transformer-based sequential modeling has emerged as a new paradigm for advancing large-scale recommendation research~\citep{kang2018self, trans4rec, hstu, si2024twin} in the era of scaled-GPU computation, Particularly, traditional deep-learning based recommendation models (DLRM) have focused on feature interactions~\citep{wang2017deepcrossnetwork} with carefully engineered human features. While effective, these models do not scale efficiently with increased compute~\citep{hstu, wang_scaling_2025} on more feature interactions or additional layers.
In contrast, transformer-based sequential modeling, which emphasizes end-to-end learning from raw user behavior sequences, can jointly capture both long-term preferences and short-term intent~\citep{uih} and exhibit favorable scaling laws with compute: model performance improves with longer sequences, denser computation in attention layers, and increased depth in stacked attention layers.

A prominent line of research in this area is the development of Hierarchical Sequential Transduction Units (HSTU) \citep{hstu}, which introduces a customized transformer-style architecture designed to efficiently learn user interests directly from raw sequential data. HSTU is notable for being the first to demonstrate favorable scaling properties with a transformer-like approach specifically tailored for recommendation systems. The sequential modeling paradigm has since been widely adopted and further advanced by leading industry practitioners, including Douyin~\citep{10kdouyin, longer}, Meituan~\citep{Han_2025}, Alibaba~\citep{wang_scaling_2025}, Xiaohongshu~\citep{huang_towards_2025}, Meta~\citep{hstu} and Linkedin~\citep{hertel_efficient_2024}, each contributing their own architectural innovations. This broad adoption across major platforms underscores the effectiveness and impact of sequential modeling in large-scale recommendation systems.

However, transformer-based recommendation models including HSTU suffer from a complexity of $\mathcal{O}(L^2)$ due to the self-attention mechanism, where $L$ denotes the length of the user history sequence. This quadratic scaling quickly becomes impractical~\citep{zaheer2020big,vaswani2017attention} when attempting to model user histories containing 
$\mathcal{O}(10\text{k})$ to 
$\mathcal{O}(100\text{k})$ events, especially in environments where it is standard to serve
billions of recommendations daily with sub-second latency. 
To mitigate the quadratic computational bottleneck, previous approaches from industry leaders have primarily adopted \emph{cross attention}~\citep{longer,10kdouyin}, using only ranking candidates or truncated user histories as queries instead of \emph{self-attention}, which considers the entire user history. Alternatively, some methods restrict themselves to shallow architectures, employing only 2–4 attention layers \citep{10kdouyin}. These strategies fundamentally diverge from practices in large language models (LLMs). While these techniques substantially reduce computational complexity, they may forgo the benefits of powerful self-attention mechanisms and deeper model architectures. As demonstrated in our experiments (see Table ~\ref{tab:model_comparison}, \ref{tab:cross_attn}), self-attention remains superior to cross attention in industrial settings, especially in terms of enabling stacked layers or scaled up computation.
This distinction represents a key research finding and highlights a major difference of our work from prior solutions: rather than eliminating self-attention, our work focuses on efficiently harnessing its advantages from model and system co-optimizations inspired by DeepSeek-V2~\citep{deepseekv2}.

\begin{figure*}
    \centerline{\hfill\includegraphics[width=1.0\columnwidth]{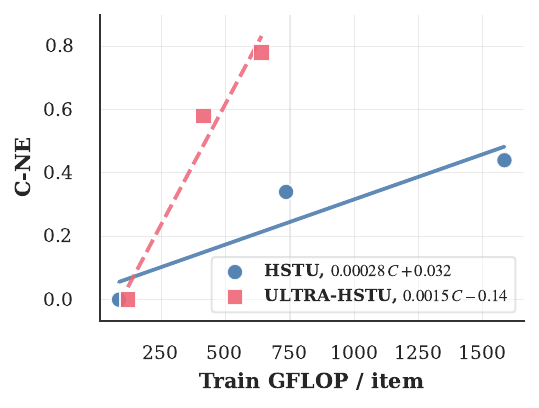}\hfill \includegraphics[width=1.0\columnwidth]{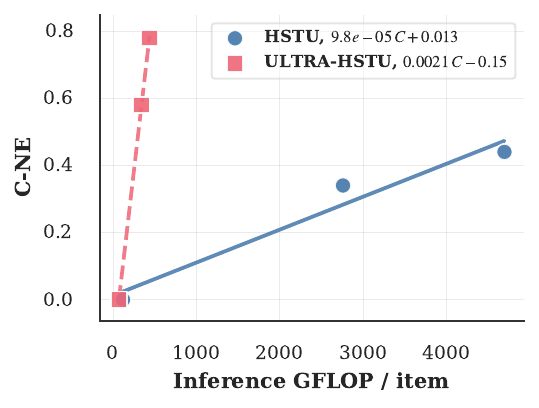}\hfill}
    \caption{Overall Performance: Scaling performance with respect to train (left) and inference (right) FLOP. 
   Compared to vanilla HSTU, ULTRA-HSTU has more than 5.3$\times$ training scaling efficiency and 21.4$\times$ inference scaling efficiency.}
    \label{full_scaling_linear}
\end{figure*}

To bend the \emph{scaling efficiency} of scaled ultra-long user history modeling, we introduce ULTRA-HSTU design, the next-generation HSTU model with a comprehensive suite of detailed model and system optimizations inspired by DeepSeek-V2~\citep{deepseekv2}.  Here, \emph{scaling efficiency} is formally defined as the slope of the fitted linear regression between model performance and computational cost. Under fixed input sequence configurations, our optimizations achieve more than $21\times$ inference scaling efficiency and $5\times$ training scaling efficiency relative to the original HSTU architecture~\citep{hstu}. This advancement effectively bends the scaling curve of recommendation systems, enabling model quality to accelerate substantially when scaling computational resources (see Figure~\ref{full_scaling_linear}).

To validate the proposed solution, we deployed ULTRA-HSTU, with 18 layers of self-attention over $16k$ user behavior sequences trained with multiple hundreds of H100 GPUs, in a large-scale production environment serving billions of users. It achieves substantial consumption and engagement improvements ranging from $4\%$ to $8\%$ as well as a $0.217\%$ uplift in topline metrics. This demonstrates both the scaling potential of sequential modeling in the recommendation domain and the effectiveness of our solutions. To our knowledge, ULTRA-HSTU stands as one of the largest sequential models ever deployed in industry, demonstrating substantially improved scaling efficiency. 
We summarize the technical innovations of ULTRA-HSTU as follows.

    \textbf{Input sequence optimizations}. We introduce two complementary designs to optimize original HSTU’s input sequence processing. First, we effectively merge item and action representations in sequence designs and enhance this simplified design with heterogeneous action encodings. Second, to mitigate inefficiency caused by cross-rank sequence-length imbalance in synchronous distributed training, we propose Load-Balanced Stochastic Length, enforcing a per-rank compute-load constraint during stochastic length sampling to reduce stragglers and improve training throughput by 15\%.
    
    \textbf{Model-system co-design for extremely efficient attention}.
    We provide an end-to-end model–system co-design that makes self-attention in HSTU viable for ultra-long user interaction history modeling in production by eliminating the common quadratic and kernel overheads. On the modeling side, we introduce a semi-local attention (SLA) mechanism tailored to the structure of user behavior sequences, achieving efficient linear sparse attention with $\mathcal{O}((K_1 + K_2) \cdot L)$ complexity without sacrificing model quality, where $K_1$ and $K_2$ 
    are local and global window sizes respectively. 
    We show that SLA improves the inference scaling efficiency more than $5\times$ compared to baseline models.
On the system side, we pair SLA with fine-tuned, hardware-aware optimizations that remove practical bottlenecks and improve hardware utilization in both training and inference. 
    We co-design a recsys-tailored mixed-precision framework spanning 16/8/4-bit formats: we keep most of the operations in BF16 for stability, accelerate the dominant GEMM computations with FP8, and reduce inference communication traffic with INT4 embedding quantization.
    We further extend FlashAttention V3~\citep{shah2024flashattention} ideas, building custom SLA kernels that handle HSTU’s SiLU-based attention and non-standard masks, and tuning them for heterogeneous GPU architectures (NVIDIA H100 and AMD MI300) to sustain high GPU utilization. 
    We also introduce memory saving optimizations with minimal efficiency overhead that significantly cut HBM footprint, enabling ultra-long sequence training. 
    Together, these co-designed components enable 70\% training and 50\% inference throughput gains over the same model without these system optimizations. Note that to maximize end-to-end performance, we focus on end-to-end throughput (how fast it finishes training/inference for a fixed number of examples) given the same model performance, instead of purely optimizing GPU utilization.  

    \textbf{Dynamic topological model designs.} 
   Scalability in recommendation models extend beyond sequence length, vertical scaling through stacking additional layers yields additional benefits, particularly by increasing capacity via residual connections~\citep{resnet}. However, naively stacking HSTU with SLA incurs a  cost of $\mathcal{O}(D L)$ where $D$ is the depth of the model. Building on the insight that different user signals yield different predictive values, we propose two novel topological designs to focus computation on most important signals. Specifically, we propose 1) Attention Truncation, which runs the first $N_1$ layers on the full sequence, then selects a shorter valuable segment and applies additional $N_2$ layers only on that segment; and 2) Mixture of Transducers (MoT), which processes heterogeneous behavioral signals as multiple sequences with separate transducers and fuses their representations, enabling targeted capacity/compute allocation to high-value signals rather than forcing everything to compete in one timeline. In our experiments, both topological designs significantly improve performance and efficiency tradeoff that further upscales our scaling capabilities.

\section{Related Work}
Traditional industrial-scale recommendation models usually follow the deep learning recommendation model framework \citep{ 
naumov2019deep,
mudigere2022software} that focuses modeling user and item feature interactions. 
The past few years have witnessed a paradigm shift in how large-scale recommendation models are trained in industry. Rather than relying on cross-user-item features, much of the recent progress in recommendation systems has been driven by learning from user interaction histories. Deep interest networks (DIN) \citep{zhou2018deep} were one of the classic short-sequence learning methods. SASRecs \citep{kang2018self} is a traditional transformer implementation for recommendations. HSTU \citep{hstu} was later proposed to perform better than traditional transformer-based models in recommendations with target-aware predictions. By capturing the implicit and explicit information learnt from raw user interaction history, HSTU has removed its dependency on human-crafted user-item features and exhibits a favorable scaling law.

Building on this line of work, our paper focuses on further improving the scaling behavior, aiming to achieve better models at reduced computational cost. Closely related to our work is research focused on improving the training and inference efficiency of sequential models. With notable breakthroughs in native sparse attention (NSA) \citep{nsa}, linear sparse attention has become the focus to deploy scalable big models \citep{beltagy2020longformer}. In addition to exploring sparse attention, Stacked Target-to-History Cross Attention (STCA) \citep{10kdouyin} was proposed to only run query focused on ranking targets, which significantly reduces the model complexity but introduces performance regressions due to the simplified attention mechanism without self-attention. Although STCA achieves linear complexity, more expensive pre-attention projections were introduced in STCA to improve the performance with a big computational overhead, making it less effective in capturing information from shorter sequences (see Table \ref{kuairand_oss_exp_res} for details).

\begin{figure*}[ht]
  \begin{center}
    \centerline{\includegraphics[height=8cm]{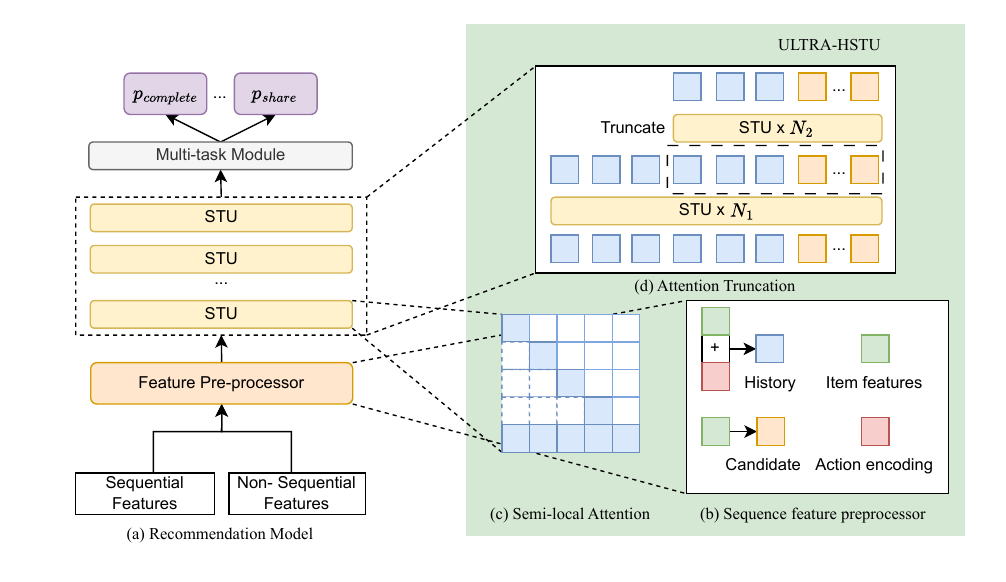}}
    \caption{Model design overview: (a) General recommendation model design. (b) Input sequence optimizations with action-aware designs (c) Semi-local attention mask with linear complexity (d) Attention truncation for dynamic topological designs.
    }
    \label{big_pic}
  \end{center}
\end{figure*}

\section{Background}

As illustrated in Figure \ref{big_pic} (a), 
a typical recommender system takes input features and trains on multi-task classification problems.
Formally, it learns a multi-task model $\mathcal{M}$ to output a probability $\hat y_k = \mathcal{M} (X, x_j) \in [0,1]$ for a candidate $x_j$ across different prediction tasks $y_k$ (e.g., like, video completion, comment), and rank the candidates based on predicted scores. Here $X$ is the input features of user. We optimize the model by minimizing the cross-entropy loss between predictions $\hat y_k$ and ground truth labels $y_k$ collected from logs. Throughout the paper, we use $L$ to denote the general sequence length, $X$ to denote the input features. Most generative ranking paradigm models the input $X$ as a sequence of embeddings (see context below), and leverage attention layers to learn probabilities from those sequential embeddings.

\paragraph{Input} As shown in Figure \ref{big_pic} (a), a recommender uses a feature preprocessor to transfer different input features into a sequence of embeddings, including:
    \textbf{user interaction history (UIH) sequence} records a sequence of items a specific user interacted with and the corresponding actions (e.g., like, comment, video completion, etc.) and context (e.g., timestamp).  Raw item IDs (and its multi-modal representation), action types are represented by $d$-dimensional learnable embeddings via embedding table lookups.  For user $i$, we denote its UIH as $X_i = \{I_i, A_i\}$, where the item embeddings in UIH are $I_i = \{I_{i,j}\}_{j=1}^{L_i} \in R^{L_i, d}$, and action embeddings are $A_i = \{a_{i,j}\}_{j=1}^{L_i} \in R^{L_i, d}$ and $L_i$ is the total length of user $i$'s UIH. 
    \textbf{Non-sequential features} include user-side features, such as country, user language, etc., and item-side features, such as  sparse (e.g., raw ID of item) and dense (e.g., click-through rate for this item) features. Those user-side features could be summarized into context-embeddings and put at the beginning of sequential UIH~\citep{hstu}. Those item-side features could be summarized as item-embeddings and inserted into sequences as target-side embeddings~\citep{zhang2025onetrans}.

\paragraph{Model} Given a sequence of embeddings, modern recommender leverages transformer-style models. One typical arch is Hierarchical Sequential Transduction Units (HSTU) \citep{hstu}, which shows significant wins on top of vanilla transformers in recommender systems by the following modifications: 

\begin{align}
\text{normaliation:   } &  X = \text{Norm}(Z) \label{eq:norm}\\
\text{pre-attention:   } \text{  }& U, Q, K, V = \phi_1(f_1(X)) \\
\text{attention:   } \text{  } & A = \big(\phi_2 (QK^T)\odot M \big) V \label{vanilla_attn}\\
\text{post-attention:   } \text{  } & Y = f_2\left(\text{Norm}\left(A\right) \odot U\right) \label{eq:post_attn}\\
\text{residual connection:   } \text{  } & Z = Y + Z \label{eq:residual}
\end{align}

Here $\odot$ denotes an elementwise product, $f_1$ and $f_2$ are MLPs for pre-attention and post-attention projections respectively, $\phi_1$ and $\phi_2$ are SiLU activations. In Equation \ref{vanilla_attn}, a causal mask $M$ is applied to maintain the temporal relationship between sequential items. Input embeddings $Z$ are normalized before passing to later operations, each layer connects the output from the previous layer $Y$ by standard residual connections~\citep{resnet}. 

While HSTU shows favorable scaling laws for recommendation systems, we believe the scaling curve can be further optimized via a model and system co-design approach similar to Deepseek-V2 \citep{deepseekv2} for LLM. Thus, we introduce ULTRA-HSTU on top of the original HSTU, and the ideas discussed below can be generally applied to other attention architectures for sequential recommendation.

\section{ULTRA-HSTU: Extremely Efficient High-performance Sequential Encoder}
To bend the scaling curve of vanilla HSTU, we make significant improvements over three key areas below: 1) input sequence optimization reduces the effective sequence length at the source; 2) a recommender-tailored sparse attention computation achieves linear complexity; 3) dynamic topological design further enables favorable depth scaling without paying full-sequence cost in every layer. Beyond theoretical complexity reductions, ULTRA-HSTU is model and hardware co-designed for practical efficiency in large-scale distributed training and inference recommendation settings.
Together, we present ULTRA-HSTU below with more than $21\times$ inference scaling efficiency and $5\times$ training scaling efficiency compared to vanilla HSTU. The general architecture of the model design is detailed in Figure \ref{big_pic} and we present each components in the following sections.

\subsection{Input Sequences Optimization}
We first propose an efficient action encoding method to effectively shorten the input sequence length by $2\times$ and thus improve the efficiency by $4\times$ in attention computation.
Recall that the vanilla HSTU \citep{hstu} interleaves items and actions, formulating the sequence input for user $i$ as $\{I_{i,1}, a_{i,1}, I_{i,2}, a_{i,2}, \dots, I_{i,L_i}, a_{i,L_i}\}$. While this generally supports both retrieval and ranking stages, it causes the sequence in ranking to double the actual UIH length. Directly combing actions and items may leak the action information of candidates to be predicted. Thus we mask the action embeddings in all candidate positions to be $a_{i,j} = 0_d$ if $j$ is a candidate to be ranked in recommender systems. We explored different ways of combining action and item embeddings and choose to simply add the embeddings of items and actions, formulating the sequential input of user $i$ as $X_{i} = \{x_{i,j}\}_{j=1}^{L_i}$, where $x_{i,j} = I_{i,j} + a_{i,j}$. 
We hypothesis that gradient can be more easily passed through action encodings in this method. Further, we enhance the construction of action embeddings in ULTRA-HSTU by exploring heterogeneous action encodings, from both implicit and explicit signals and side information through user contextual features. 
Importantly, this design reduces the sequence length to be half as the UIH designed in the vanilla HSTU \citep{hstu} without sacrificing model quality, allowing ULTRA-HSTU to achieve substantial performance improvements while preserving its scalability.

We further designed load-balanced stochastic length algorithm to improve training throughput by $15\%$. In \citep{hstu}, Stochastic Length (SL) randomly selects users and samples their history sequences to a predefined threshold length of $L^{\frac{\alpha}{2}}$ during the training stage, where $\alpha \in (1, 2]$ is a tunable hyper-parameter of SL. This reduces the computation complexity from $\mathcal{O}(L^2)$ to $\mathcal{O}(L^\alpha)$ during training and is proved to be able to generalize in inference with the full sequence length. Similar ideas have been adapted in other papers \citep{10kdouyin} as well. However, in a distributed training environment, this sampling process occurs independently on each rank, leading to significant variations in the input and output load (represented by the sum of user sequence lengths) on each rank.
Such load imbalances could greatly reduce training efficiency in our synchronous distributed training framework.
We propose load-balanced stochastic length (LBSL) algorithm, a variant of SL that explicitly controls per-rank compute to reduce stragglers.
Define the load of a rank in a batch as $\sum_{u\in \text{rank}}n_u^\gamma$, where $n_u$ is the sequence length for request $u$ and $\gamma$ captures the superlinear cost of HSTU ($\gamma \in (1, 2)$). Load balance is defined as the ratio between the maximum and minimum rank loads within a world size. 
LBSL runs in three stages: it first performs a short warmup using standard stochastic length to estimate a global target load $l$; then performs constrained sampling that adaptively selects an unsampled set so each rank’s realized load matches $l$ as closely as possible, while preserving SL’s bias toward leaving shorter sequences unsampled via weights $p_u$ and weighted sampling-without-replacement plus a greedy fill; finally, it applies periodic recalibration of 
$l$ on a configurable interval to track slow shifts in the production length distribution. When recalibrated every batch, LBSL matches standard SL’s average load but redistributes sampling across ranks (sampling more on heavy-load ranks and less on light-load ranks) to reduce stragglers without harming quality.
Details are in Algorithm~\ref{alg:lbsl} in Appendix.

\subsection{Model-System Co-Design for Efficiency}
\subsubsection{Semi-Local Attention Design}

We introduce a novel sparse attention mechanism called Semi-Local Attention (SLA) which achieves linear complexity in attention computation, significantly improving the inference scaling efficiency of ULTRA-HSTU by $5\times$.
Recall in Equation \ref{vanilla_attn} that vanilla HSTU models \citep{hstu} calculate full causal self-attention mask with the following formula, incurring quadratic cost when model scales the sequence length. 
\begin{equation}
    A(X) = \phi_2 \Big(Q(X) K(X)^T \Big) \odot M V(X),
\end{equation}
where $M \in R^{L\times L}$ is a causal attention mask with $M_{i,j} = 1$ only when $j \leq L - i$, $\phi_2$ is SiLU in HSTU.

\begin{figure}[ht]
  \begin{center}
    \centerline{\includegraphics[width=0.8\columnwidth]{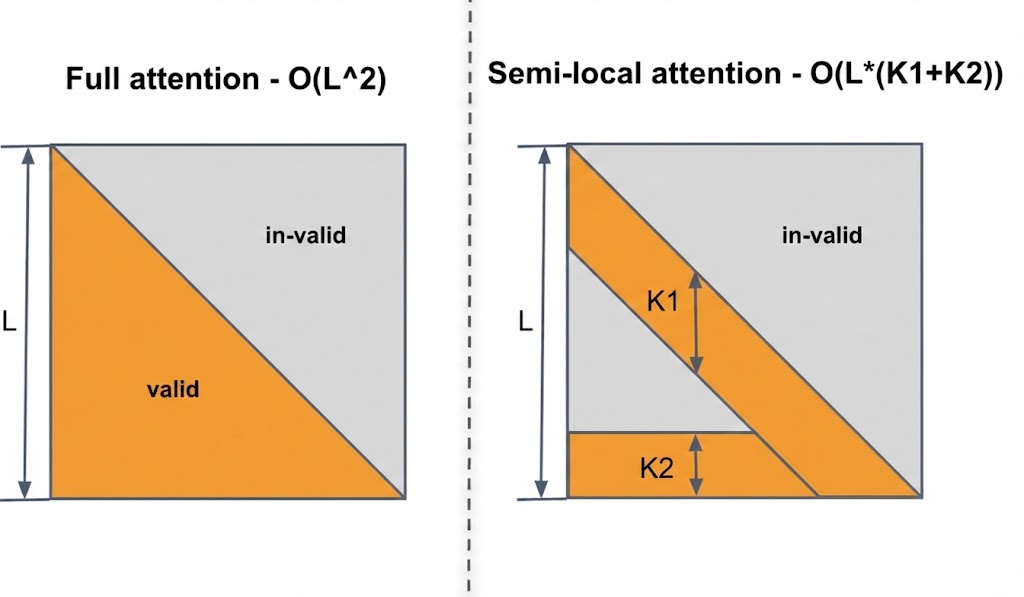}}
    \caption{Attentions masks. Left plot: full causal self-attention masks. Right plot: Semi-local attention masks.
    }
    \label{sla}
  \end{center}
\end{figure}
In large-scale recommender systems, the length of UIH quickly accumulates and goes beyond $10k$, leading to an undeployable situation in real-world ranking. Motivated by the intrinsic sparse and dynamic attention in both LLM \citep{nsa} and Recsys, we develop a semi-local attention mechanism that leverages its nature of sparsity, with a focus on both long-term and local patterns. 
We define two hyper-parameters, local window size $K_1$ and global window size $K_2$. Local window size controls the window length of local pattern that will be counted into the attention mask, and global window size focuses on the latest UIH attention patterns, capturing users' long-term interests. 
The resulting attention mask is defined as the following in semi-local attention (see Figure \ref{sla} for illustration):

\begin{equation}
    M_{i,j} = \begin{cases}
    1 & \text{if } L - K_1 \leq i + j \leq L   \\
    1 & \text{if } j \leq K_2 \text{ and } j \leq L-i \\
    0 & \text{otherwise}
\end{cases}
\end{equation}
With this design, the resulting computational complexity in attention is reduced to linear as $\mathcal{O}((K_1 + K_2) \cdot L)$, significantly improving model efficiency when sequence length $L$ scales beyond $10k$ in large-scale recommender systems.
In contrast to native sparse attention (NSA) from DeepSeek \citep{nsa} where only local window is applied, we will show in Section \ref{exp} that both the design of local and global windows are necessary, which was especially pronounced in recommenders where users' long-term behaviors are critical. 

\subsubsection{System optimizations} \label{sec:system_opt}
\paragraph{Mixed-Precision Training and Inference}

\begin{figure}[ht]
  \begin{center}
    \centerline{\includegraphics[width=1.0\columnwidth]{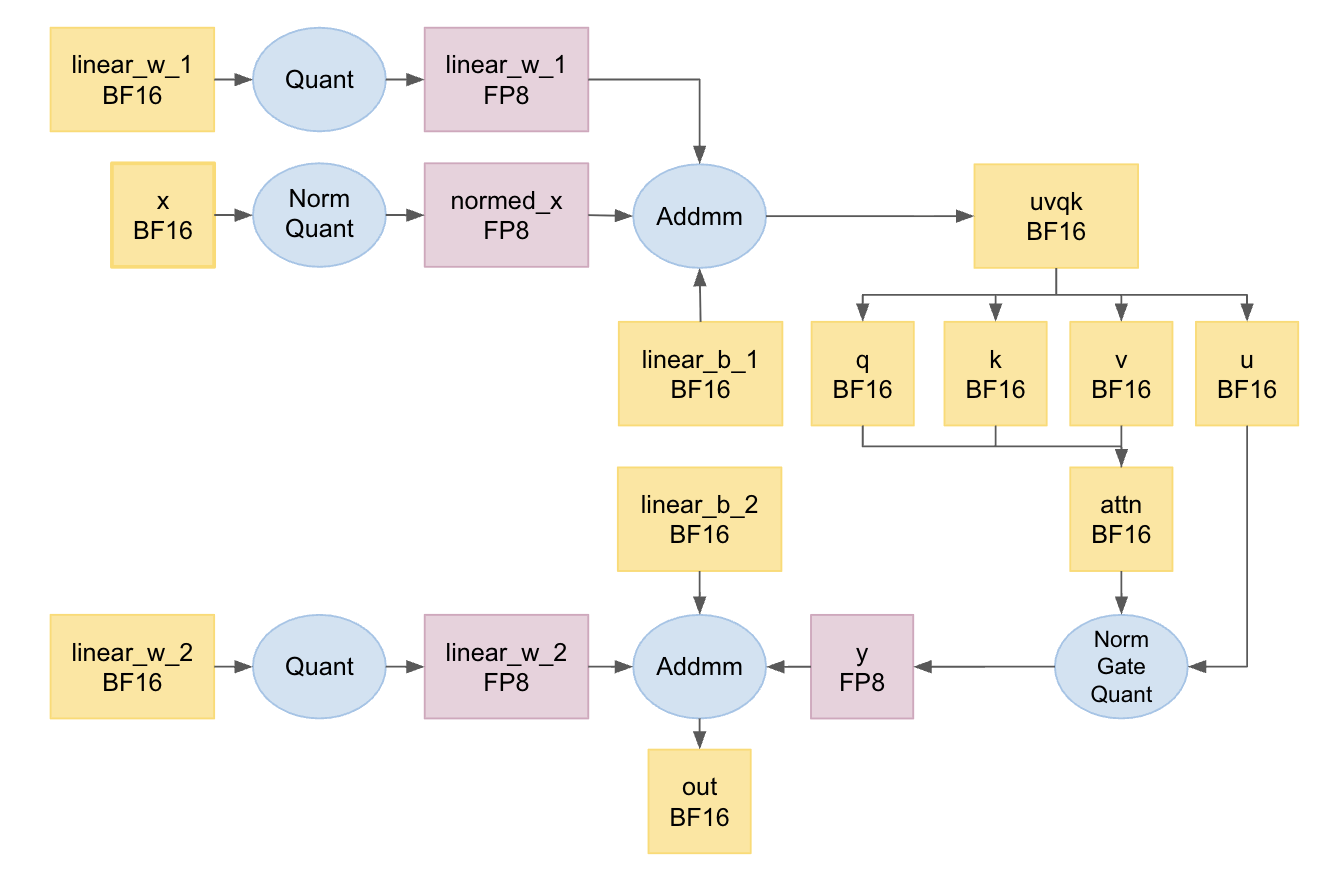}}
    \caption{Mixed precision computation framework. We fuse the scaling/quantization steps with the preceding kernels.
    }
    \label{fp8}
  \end{center}
\end{figure}

Large-scale recommendation models are bottlenecked by a combination of dense compute including General Matrix Multiplications (GEMMs) and data movement, especially embedding lookup and host-to-device transfer in serving. To make ULTRA-HSTU efficient end-to-end, we co-design a recsys-tailored mixed-precision framework spanning \textbf{16/8/4-bit} formats: we keep most of the operations in BF16 for stability, accelerate the dominant GEMM computations with FP8, and reduce inference communication traffic with INT4 embedding quantization.
In both offline and online experiments, this mixed-precision stack yields 10\% training and 40\% serving throughout improvement while preserving model accuracy.
We develop a customized FP8 stack (shown in Figure \ref{fp8}) for HSTU that targets two practical bottlenecks simultaneously: improving Tensor Core utilization on NVIDIA H100 to raise achieved TFLOP/s on the dense compute, and reducing the overhead of FP8 quantization/scaling that can otherwise become memory-bandwidth bound. Each HSTU layer contains two GEMMs: a projection that maps input embeddings $X$ into the $U,V,Q,K$ tensors prior to attention, and a post-attention projection that transforms the normalized and gated attention output to produce the layer output. We execute both GEMMs in FP8, while preserving all remaining operations in BF16, improving throughput without sacrificing numerical robustness.
Simply switching GEMMs to FP8 is inefficient: naive FP8 pipelines need an additional operation on scaling/quantization and layout preparation which can offset the expected speedup. To ensure FP8 accelerates end-to-end training/inference, we develop fused kernels that combine row-wise scaling computation and quantization with the preceding layer-normalization kernels (Equations~\ref{eq:norm} and \ref{eq:post_attn}) for jagged embeddings, eliminating extra passes over memory and reducing quantization overhead. 
We further develop high-performance Triton FP8 GEMM kernels specifically for the post-attention projection. In this path, the projection output must be accumulated with a 2D residual tensor (Equation~\ref{eq:residual}), so we fuse this residual accumulation directly into the GEMM epilogue. This is not efficiently supported by PyTorch GEMM kernels, which typically assume a 1D bias vector. Our Triton FP8 kernel natively supports 2D bias, while leveraging persistent scheduling, TMA, warp specialization, and epilogue pipelining to sustain high throughput without excessive register pressure.
In addition to FP8 GEMM, our mixed precision framework incorporates 4-bit quantization for embedding movement during serving. Further details are provided in Appendix~\ref{other_system}.

\begin{figure*}[htbp]
  \begin{center}
    \centerline{\hfill\includegraphics[width=0.6\columnwidth]{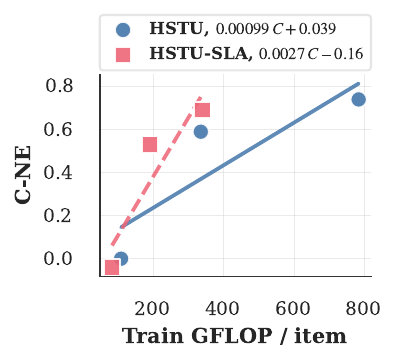}\hfill \includegraphics[width=0.6\columnwidth]{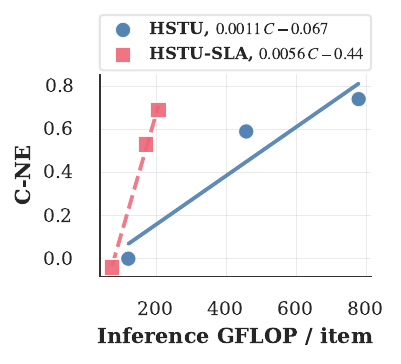} \hfill
    \includegraphics[width=0.6\columnwidth]{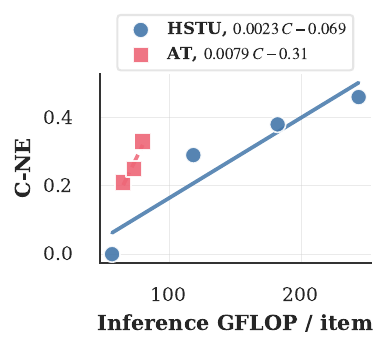}\hfill }
    \caption{Ablation study on scaling: training (left) and inference (middle) of SLA, inference (right) of attention truncation (AT).}
    \label{fig:sla_scaling}
  \end{center}
\end{figure*}

\paragraph{Efficient SLA Kernels for Heterogenous Hardware}
Attention operations are a bottleneck for HSTU. In~\citep{hstu}, the kernel is implemented with Triton~\citep{tillet2019triton} using the FlashAttention V2 algorithm~\citep{dao2024flashattention}. 
We improve this baseline by adopting the FlashAttention-V3 algorithmic design~\citep{shah2024flashattention} to aggressively overlap data movement and compute, while customizing the kernel to HSTU's non-standard attention (pointwise SiLU activation and SLA masking). We implemented this design on both NVIDIA H100 and AMD MI300x, allowing heterogeneous service and delivery \textbf{\( 2\times\)} speedup over the FlashAttention-V2 baseline on both platforms.
On NVIDIA H100, we implement CUDA kernel families for both full and semi-local HSTU attention using FlashAttention-3-style pipelining. We also implement an analogous kernel on AMD MI300x via Composable Kernel~\citep{amd2025composablekernel}. Since MI300x lacks H100 features leveraged by FlashAttention-3 (e.g., TMA and warp-specialized async execution), we introduce MI300x-native optimizations: XCD-aware scheduling to exploit the 8-chiplet topology, LDS layouts to reduce shared-memory bank conflicts, and explicit VMEM/MFMA interleaving via scheduling barriers---and obtain a \textbf{\( 2\times\)} speedup over the Triton kernel baseline.

\paragraph{Memory Saving with Minimal Overhead}
The standard attention implementation incurs high GPU memory pressure in the forward pass, which becomes a primary bottleneck for ultra-long sequence training. We carefully designed the following optimizations to save memory and preserve training efficiency. 
First, we introduce selective activation rematerialization specialized for ULTRA-HSTU. Specifically, we skip saving six large forward tensors and reconstruct them in backward with minimal recompute, including reusing saved layer-norm statistics for normed $X$, rerunning GEMM to recover $U, Q, K, V$, and computing the intermediate  $Y$ inside the fused gated normalization kernel. This is substantially lighter than generic checkpointing, with only 
5\% overhead compared to the baseline without any activation recomputation.  The detailed algorithm is in Listing~\ref{lst:hstu_full_layer} in Appendix.
Second, we remove that overhead by eliminating gradient concatenation for $dU, dQ, dK, dV$, cutting memory traffic and kernel overhead in backward. Overall, ULTRA-HSTU achieves about 67\% per-layer memory reduction with no regression in efficiency. With a 512 embedding dimension size, 256 batch size, and $3k$ sequence length and BF16 data type, the technique reduces the HBM memory usage per layer from 7GB to 2.3GB.
Thrid, we employ a fully jagged tensor implementation for end-to-end training, eliminating the need for padding to dense tensors and significantly reducing memory usage. 
We present detailed efficiency benchmark experiments in Appendix \ref{eff_bench}.

\subsection{Dynamic Topological Design}\label{topology}

In addition to scaling sequence length, depth scaling are crucial for model performance. Naively stacking ULTRA-HSTU layers with SLA each processing the full sequences incurs a computational cost of $\mathcal{O}(D L)$ where $D$ is the model depth. In real-world applications, when $L$ scales to $10k$ sequence lengths, stacking more and more model layers introduces significant training, memory and inference costs even when linear sparse attention are already in place. However, the necessity to be able to handle millions of requests in large-scale recommender systems within milliseconds remains the same. One natural question is that do we really need attention on full sequences in every layer when we stack more and more layers? Motivated by this, we propose 
two efficient topological designs below to further improve the scaling law of ULTRA-HSTU.

\textbf{Attention Truncation}. Motivated by the importance of users' most recent interaction history,
after stacking $N_1$ layers of HSTU with full sequence length $L$, we propose to select a segment of length $L^\prime$ from the full sequence, and stack another $N_2$ layers of HSTU on the selected UIH segment only. Many methods can be applied to select this UIH segment, including 1) truncating latest UIH of length $L^\prime$; 2) applying Stochastic Length (SL) \citep{hstu} on top of the first SL to select sequence length of $L^\prime$; 3) inserting a compression modules after the first $N_1$ layers to compress the full sequence to length $L^\prime$. In practice, we found that simply truncating the latest UIH segment gives the best model performance (See Figure \ref{big_pic} (d)).

\textbf{Mixture of Transducers}. Recommendation models inherently process multiple input sequences, as user engagement signals from various sources and types are typically recorded separately. Aggregating all user signals into a single input sequence for a unified encoder compresses heterogeneous user interactions into one timeline can dilute sparse, high-value engagements among dense, implicit signals and force all signals to compete for limited sequence capacity.
To address this challenge, we introduce the Mixture of Transducers (MoT) paradigm. MoT processes multiple distinct input sequences through separate transducers and subsequently fuses the learned user embeddings. This approach enables the model to capture different types of user behaviors over varying time spans, resulting in a more granular and effective representation of diverse and sparse engagement patterns. Crucially, MoT allows for flexible allocation of computational resources across input sequences. For instance, the model can assign deeper layers and greater capacity to high-value sequences, while reducing resources for well-understood or less critical sequences. This targeted allocation of computation budget ensures that the model focuses its capacity on the most meaningful user interactions, thereby improving overall recommendation quality and efficiency tradeoff.

Both of the proposed topological designs achieve significantly better model quality and cost tradeoff than vanilla HSTU. The design of Attention Truncation and MoT are compatible with each other and can be combined into one model. In real applications, the choice of the topological designs depends on the most concerning metrics (efficiency or model quality) in the system. In our experimental setups reported below (Section \ref{exp}), we choose attention truncation due to its simplicity and powerful model quality and efficiency tradeoff. We defer the study of MoT to Appendix \ref{app_topology}.

\section{Experimental Results}\label{exp}

Throughout this section, we measure model quality by normalized entropy (NE), defined as the model's cross-entropy divided by the cross-entropy by purely making predictions based on mean frequency of positive labels~\citep{he2014practical}. 
Formally, NE is defined in the following equation:
\begin{equation}\label{ne}
    \text{NE} = \frac{ -\frac{1}{N} \sum_{i=1}^N (y_i \log p_i + (1-y_i) \log(1-p_i)) }{ 
    -p \log p - (1-p) \log (1-p)},
\end{equation}
where $N$ is the number of training examples, $y_i \in \{0, 1\}$ is the
label for example $i$, $p_i$ is the model prediction for example $i$, and
$p =\sum^N_{i=1} y_i / N$. The model is better if the NE is lower.
Specifically, we measure the NE improvement of a consumption task (e.g., video view complete) and an engagement task (e.g., share), denoted as \textbf{C-NE} and \textbf{E-NE}.  Here, we choose to report NE to follow the original HSTU paper~\citep{hstu} and the best practices internally ~\citep{he2014practical}. Based on our experiences and experiments, AUC and other metrics move in a consistent direction (better or worse) and at similar scales with NE. We omit reporting them due to space limitations.  

We evaluate our model against several strong baselines, categorized by their ability to model short- or long-range user behaviors. Short-sequence methods include DIN \citep{zhou2018deep} and SASRecs \citep{kang2018self}. Long-sequence methods include vanilla HSTU \citep{hstu}, STCA \citep{10kdouyin}. In addition, we also compare our methods to an internally optimized transformer which has additional projections and normalization to stabilize training and avoid unexpected metric regressions in the classic transformers for recommender systems.  

\subsection{Industrial Dataset Benchmark}

\subsubsection{Dataset}
We first report our model's performance using industry-scale production datasets sourced from internal, large-scale, real-world recommender systems. The dataset consists of a subset of online user interaction histories, totaling over 6 billion samples, each featuring ultra-long user interaction sequences with lengths ranging from $3,072$ to $16,384$ events. To ensure temporal consistency and prevent future data leakage, we employ a chronological data split: the initial $85\%$ of the data is allocated for training, while the remaining $15\%$ is reserved for evaluation. 

Please note that, we use LBSL for all the experiments on the industrial dataset. For example, when raw sequence length in inference is $16,384$, training sequence length is around $4,400$ after applying LBSL. Based on~\citep{hstu}, comparing with models trained with full sequence length, LBSL has minimum NE differences while achieving significant training speed up. This is the reason that our inference FLOP per example is higher than training FLOP per example. For more detailed sequence length comparisons during training and inference, please see Table \ref{train-eval-seq-len} in Appendix.

\subsubsection{Overall Performance}  Table~\ref{tab:model_comparison} shows the results on all methods with sequence length capped at $3,072$. We tune the model depth/parameters to allow approximately matched FLOP on all methods.
We observe that ULTRA-HSTU significantly outperforms all other methods. STCA which heavily relies on cross attention for linear complexity performs worse than ULTRA-HSTU due to lacking the power from self-attention. Note that model is worse when $\Delta$ NE is positive. Based on our experience, an improvement in the range of $0.03\%-0.05\%$ is regarded as significant and can lead to substantial gains in online metrics.

\begin{table}[ht]
\textbf{}\begin{sc}
\small
\centering
\begin{tabular}{lccc}
\toprule
{Model} & {$\Delta$ C-NE} & {$\Delta$ E-NE} \\
\midrule
ULTRA-HSTU        & $0\%$  & $0\%$ \\
HSTU            & $+0.43\%$ & $+0.04\%$ \\
STCA            & $+0.94\%$ & $+0.74\%$ \\ 
Transformer    & $+0.57\%$ & $+0.59\%$ \\
DIN  & $+1.41\%$ & $+1.91\%$ \\
SASRecs & $+1.12\%$ & $+1.28\%$ \\
\bottomrule
\end{tabular}
\caption{Model performance on Industrial datasets.}
\label{tab:model_comparison}
\end{sc}
\end{table}

\subsubsection{Scaling Law}
To analyze the scaling behavior, we fix the input sequence designs for both ULTRA-HSTU and vanilla HSTU and report the model performance on C-NE and TFLOP comparisons with improved model architecture and topological designs.
We vary the number of modeling layers from 6 to 18 and sequence length $L \in \{3072, 8192, 16384\}$ while fixing model 
dimensionality at $d=512$. Table \ref{tab:internal_3816k} reports detailed model performance and TFLOP numbers. As sequence length and the number of layers increase, we saw significantly improved efficiency and C-NE metrics from ULTRA-HSTU.
In Figure~\ref{full_scaling_linear}, we present the C-NE gain relative to the baseline model as a linear regression of computational cost for both ULTRA-HSTU and vanilla-HSTU. Notably, by comparing the slope of the fitted linear function, ULTRA-HSTU demonstrates a remarkable advancement in scaling efficiency, achieving a \textbf{5.3$\times$ improvement in training scaling efficiency and an outstanding 21.4$\times$ enhancement in inference scaling efficiency}.
\begin{table*}[ht]
\centering
\begin{tabular}{lcccccc}
\toprule
Model & Sequence length & $\#$ layers & $\Delta$ C-NE & Training TFLOP & Inference TFLOP \\
    \midrule
\multirow{3}{*}{HSTU} 
    & 3072 & 6 & $0.0\%$ & 0.085 & 0.118 \\
    & 8192 & 11 & $-0.34\%$  & 0.735 & 2.756 \\
    & 16384 & 10 & $-0.44\%$  & 1.584 & 4.692 \\
\midrule
\multirow{3}{*}{ULTRA-HSTU} 
    & 3072 & 14 & $-0.00\%$  & 0.119 ($\uparrow 40.0\%$) & 0.070  ($\downarrow 40.7\%$) \\
    & 8192 & 18 & $-0.58\%$  & 0.414 ($\downarrow 43.7\%$)  & 0.337 ($\downarrow 87.8\%$)\\
    & 16384 & 18 & $-0.78\%$  & 0.639 ($\downarrow 59.7\%$) & 0.436 ($\downarrow 90.7\%$) \\
\bottomrule
\end{tabular}
\caption{Scaling of ULTRA-HSTU on industrial datasets.}
\label{tab:internal_3816k}
\end{table*}

\subsection{Open source dataset} 
Methods such as ULTRA-HSTU and STCA \citep{10kdouyin} are designed for industrial-scale recommenders where user histories span tens of thousands of interactions. To demonstrate the general applicability of our method beyond these extreme-length settings, we evaluate on public open-source benchmarks on KuaiRand\footnote{https://kuairand.com/} with much shorter sequences of length 256. Table~\ref{kuairand_oss_exp_res} shows that our approach still achieves the best NE at the lowest computational cost in both training and inference stages even under short-sequence scenarios.
STCA struggles to adapt to shorter sequences due to the expensive pre-attention computational overhead. 

\begin{table}[ht]
\begin{sc}
\small
\centering
\begin{tabular}{lcccccc}
  \toprule
  Model & Training & Inference & NE \\
    & TFLOP & TFLOP & \\
  \midrule
  STCA & $626.49$ & $208.75$  & $0.8689$ \\
  Transformer & $802.39$ & $267.46$ & $0.8688$ \\
  SASRec & $550.92$ & $176.51$   & $0.8804$ \\
  DIN & $505.08$ & $168.36$ & $0.8685$ \\
  HSTU & $617.78$ & $198.80$   & $0.8676$ \\
  \textbf{ULTRA-HSTU} & $\textbf{504.41}$ & $\textbf{166.41}$ & $\textbf{0.8626}$ \\
  \bottomrule
\end{tabular}
  \centering\caption{Comparisons on KuaiRand benchmark.}
  \label{kuairand_oss_exp_res}
\end{sc}
\end{table}

\subsection{Ablations on Scaling Studies}\label{ablation_exp}
We first briefly mention the impact from input sequence optimizations and then ablate the scaling efficiency of SLA and attention truncation by fixing the input sequence designs for both vanilla HSTU and ULTRA-HSTU.
A detailed description of our approach for analyzing scaling laws is in Appendix~\ref{app:power-scaling-law-of-ultra-hstu}.

\subsubsection{Input sequence optimization} Removing item-action interleaving in input sequence design shrinks sequence length by half and significantly reduces training FLOP by $32.5\%$ and inference FLOP by $63.5\%$ given a UIH sequence with length $3,072$. In the meantime, heterogeneous construction of action embeddings brings $0.45\%$ C-NE gain compared to baselines.
LBSL achieves $15\%$ speedup in the world size of $512$, demonstrating its effectiveness in accelerating the training of large-scale sequential models.

\subsubsection{Semi-Local Attention (SLA)}
In Figure~\ref{fig:sla_scaling} we plot the C-NE vs the total FLOP with or without SLA enabled. The SLA enabled model achieves significantly improved scaling compared to the vanilla HSTU, with {2.7$\times$ training scaling efficiency and 5.1$\times$ inference scaling efficiency}.
We note that both local window size $K_1$ and global window size $K_2$ are necessary in the design of SLA, which significantly differs from NSA \citep{nsa} where only local sliding window is enabled.
Moreover, we find that global window size $K_2$ is more important than local window size $K_1$.
For example, if we set $K_1 = 0$ and only enable global window in SLA, we observe $0.03\%$ C-NE regression. If we set $K_2 = 0$ and only enable local window in SLA, we observe $0.35\%$ C-NE regression.

\subsubsection{Dynamic Topological Design} 

When stacking more layers in vanilla HSTU, we observe significant performance gains but with unaffordable training and inference costs. 
Figure \ref{fig:sla_scaling} (right figure) plots the scaling curve comparing the performance when stacking HSTU layers with Attention Truncation (AT) and purely stacking HSTU layers with full sequences. This experiments are conducted with $n_1$ layers of sequence length $3072$ in inference (around $1110$ in training after SL) and $n_2$ layers of sequence length $512$, with $n_1 = 3, 6, 9, 12$ and $n_2 = 0, 3, 6, 9$. Attention truncation are more effective when the sequence length is longer. With LBSL enabled in model training, efficiency savings from attention truncation at sequence length around $1110$ is not significant enough, but we see much better results in inference ($3.4\times$ more effective inference scaling) with longer sequences at length $3072$.

\subsection{Online A/B Testing}
We showcase the results validated by multiple rigorous 30-day online A/B tests to evaluate the effectiveness of ULTRA-HSTU on a large-scale production video serving platform that reaches billions of users daily. We report three different kinds of online metrics: 1) online consumption metrics (C-metric), such as watch time, video completion, etc. 2) online engagement metrics (E-metric), such as likes, comments, shares, etc. 3) online topline metrics, such as number of visits, daily active users, etc. 

We upgrade the existing production model from vanilla HSTU to ULTRA-HSTU.  
The results, summarized in Table~\ref{tab:qe}, reveal substantial and highly impressive improvements across key metrics. ULTRA-HSTU delivers
significant $4.11\%$ gains in online consumption metrics and $2\%$ to $8\%$ gains in engagement metrics depending on the engagement types. Most notably, we observed remarkable enhancements in critical “top-line” metrics, which are strong indicators of overall platform health. In our system, even single-digit percentage improvements in engagement and consumption are considered major breakthroughs. Furthermore, increases of $0.05\%$ and $0.01\%$ in Top-line 1 and 2, respectively, are regarded highly significant at Meta. 
Collectively, these results provide compelling evidence for the efficacy and potential of the proposed ULTRA-HSTU approach. To the best of our knowledge, this is the largest model tested in our recommendation platform, achieving one of the largest impacts in the past few years.  

\begin{table}[htbp]
  \begin{center}
    \begin{small}
      \begin{sc}
        \begin{tabular}{ll}
          \toprule
          User Value    & Percentage of gain \\
          \midrule
          Online C-Metric 1  &  4.11\%   \\
          Online E-Metric 2 &  2.27\% \\  
          Online E-Metric 3 &  8.2\%  \\
          Online E-Metric 3  &  4.34\%   \\
          Online Top-line 1  &  0.217\%   \\
          Online Top-line 2 &  0.037\%   \\
          \bottomrule
        \end{tabular}
      \end{sc}
    \end{small}
  \end{center}
  \centering\caption{One-month online gain over production baseline.}
  \label{tab:qe}
\end{table}

\section{Conclusions}
In this work, we present ULTRA-HSTU, a novel approach of end-to-end model and system co-design that delivers substantial improvements in scaling efficiency of sequential modeling in the recommendation domain. Our contributions can be summarized as follows: 1)
as our key research findings, we show self-attention is still superior to cross-attention and scaling up computation on attention layers and sequence length continue improving model performance; 2) as our key tech innovations, we presented multiple modeling and system co-optimizations from LBSL in input processing, semi local attention, heterogeneous hardware kernel optimization with mixed precision training/inference, to dynamic model topological design, and achieved 5$\times$ training and 21$\times$ inference scaling efficiency. 3) as our key sharing with the recommendation industry, we deployed our ULTRA-HSTU, with 18 layers of self-attention over $16k$ user sequences trained on hundreds of H100 GPUs, into a large-scale production environment with significant impacts, demonstrating the promising direction of scaling up sequential models in recommendation and the effectiveness of our proposed innovations.

\section{Acknowledgement}

This work would not be possible
without contributions from the collaborators and supports from the leaderships as follows (alphabetical order): Hao Lin, Hong Yan, Jiaqi Zhai, Jie Hua, Shilin Ding, Yu He.

\bibliographystyle{plainnat}
\bibliography{paper}

\clearpage
\newpage

\appendix

\section{Selection of Topological Designs}\label{app_topology}

Both attention truncation and MoT achieve significantly better model quality and cost tradeoff than vanilla HSTU. The design of Mixture of Transducers and Attention Truncation are compatible with each other. In this section, we detail the advantage of each design. In real applications, the choice of the topological designs depend on the most concerning metrics (efficiency or model quality) in the system.

\textbf{Mixture of Transducers (MoT)}. MoT delivers significant NE gains on engagement tasks (E-NE in Table \ref{tab:model_component_comparison}) while achieving competitive training/inference FLOP savings. By decoupling heterogeneous signals into dedicated modules, MoT mitigates the signal competition that occurs when diverse input signals are constrained within a single module with limited sequence length.

Specifically, we employ two specialized HSTU modules: one for engagement events and one for consumption events, denoted as E-seq and C-seq in Table \ref{tab:model_component_comparison}. Each module processes shorter sequences compared to its single-HSTU counterpart, yet achieves richer signal representation through careful sequence composition. For instance, the dedicated engagement module, despite having a shorter sequence, captures significantly richer engagement history by mitigating competition with dense consumption signals. We further optimize computational efficiency by tailoring the compute allocation per module; the shorter sequences result in substantially lighter attention operations, yielding significant FLOP savings during both training and inference.
\begin{table}[ht]
\small
\centering
\begin{tabular}{lccccc}
\toprule
 & $\#$ layers & $\Delta$ C-NE & $\Delta$ E-NE \\
 \midrule
\multirow{3}{*}{Cross-attention} 
 & 3  & 0.00\% & 0.00\% \\
 & 6  & -0.09\% & -0.05\% \\
 & 9  & -0.14\% & -0.15\% \\
& 12  & -0.14\% & -0.17\% \\
\midrule
\multirow{4}{*}{Self-attention} 
 & 3    & 0.00\% & 0.00\% \\
 & 6    & -0.29\% & -0.27\% \\
 & 9   & -0.38\% & -0.47\% \\
 & 12 & -0.46\% & -0.65\% \\
\bottomrule
\end{tabular}
\caption{Depth scaling of cross-attention vs self-attention}
\label{tab:cross_attn}
\end{table}

\textbf{Attention Truncation}. In Table \ref{tab:hstu_vs_attntrunc_scaling_3k}, we list complimentary data of attention truncation with increasing number of layers of vanilla HSTU only or number of attention truncation layers. With deeper model layers, we observe significant NE wins in both consumption and engagement tasks, but at the cost of bigger inference FLOP burden. With attention truncation, we achieve significantly better tradeoff between model quality and computation costs. For example, comparing 3-layer vanilla HSTU stacked by another 6-layer attention truncation vs only 6-layer vanilla HSTU, attention truncation achieves on par C-NE metrics and better E-NE metrics with $3\%$ train TFLOP savings and $38\%$ inference FLOP savings.

\section{Diminishing Return of Depth Scaling in Cross-Attention}

We show in Table \ref{tab:cross_attn} that self-attention is more powerful than cross attention in terms of model depth scaling. With a sequence length around 3072, stacking more layers of cross-attention shows saturated model performance with $9$ layers, while self-attention exhibits consistently better model quality with increased number of layers.

\section{Algorithm of Load Balanced Stochastic Length}\label{lbsl_algo}

\begin{table}[ht]
\begin{sc}
\small
\centering
\begin{tabular}{lcccc}
\toprule
Raw UIH length    & 3072     & 8192  & 16384 \\
          \midrule
          Train with SL     &  1110    &  2600 & 4400 \\
          Inference without SL &  3072 &  8192 & 16384 \\
\bottomrule
\end{tabular}
\caption{Average UIH sequence length in training with SL and inference without SL.}
\label{train-eval-seq-len}
\end{sc}
\end{table}

\begin{table}[htbp]
\textbf{}\begin{sc}
\small
\centering
\begin{tabular}{lccc}
\toprule
{(m, k \& n)} & \shortstack{Bias-Fused \\ FP8} &
\shortstack{Bias-Split \\ FP8} & BF16\\
\midrule
(1M, 512)    & 414   & 236  &  292   \\
(1M, 1K)     & 734  &  468 & 410    \\
(250K, 512)  &  414    & 238  & 281   \\
(250K, 1K)   &  721   & 466  & 399   \\
\bottomrule
\end{tabular}
\caption{Benchmarking FP8 GEMM kernel efficiency with 2D bias. Performance is reported in TFLOP/s under various $m, n, k$ on NVIDIA H100. Bias-Fused FP8 uses our Triton kernel with native 2D bias support, whereas Bias-Split FP8 uses Torch FP8 GEMM plus a separate bias addition since it does not support the 2D bias case.}
\label{tab:fp8}
\end{sc}
\end{table}

Details of Load Balanced Stochastic Length (LBSL) are listed in Algorithm \ref{alg:lbsl}. With LBSL enabled in our experiments in Section \ref{exp}, the training and inference sequence length when varying raw sequence length are listed in Table \ref{train-eval-seq-len}.

\begin{table*}[ht]
\small
\centering
\begin{tabular}{lcccccc}
\toprule
Model Component & Raw Seq Length & $\Delta$ C-NE & $\Delta$ E-NE & Training TFLOP & Inference TFLOP \\
\midrule
    MoT vs 3k HSTU & 2k C-seq and 1k E-seq  & $+0.01\%$ & $-1.00\%$ & 0.138 ($\uparrow 6\%$) & 0.076 ($\downarrow 53\%$) \\
    MoT vs 16k HSTU & 10k C-seq and 3k E-seq & $+0.05\%$ & $-0.30\%$ & 0.872 ($\downarrow 45\%$) & 1.03 ($\downarrow 69\%$) \\
\bottomrule
\end{tabular}
\caption{Performance of MoT.
MoT significantly improves E-NE metrics. (C-seq denotes the consumption sequences and E-seq denotes the engagement sequences.)}
\label{tab:model_component_comparison}
\end{table*}

\begin{table*}[ht]
\small
\centering
\begin{tabular}{llcccc}
\toprule
Model & Setup & Train TFLOP & Inference TFLOP & $\Delta$ C-NE & $\Delta$ E-NE \\
\midrule
\multirow{4}{*}{Vanilla HSTU} 
 & 3-layer HSTU  & 0.0427 & 0.0561 & 0.00\% & 0.00\% \\
 & 6-layer HSTU   & 0.0851 & 0.1180 & -0.29\% & -0.27\% \\
 & 9-layer HSTU   & 0.1284 & 0.1821 & -0.38\% & -0.47\% \\
 & 12-layer HSTU  & 0.1705 & 0.2438 & -0.46\% & -0.65\% \\
\midrule
\multirow{3}{*}{Attention Truncation (AT)} 
 & 3-layer HSTU + 3-layer AT  & 0.0626 & 0.0646 & -0.21\% & -0.24\% \\
 & 3-layer HSTU + 6-layer AT & 0.0825 & 0.0729 & -0.25\% & -0.31\% \\
 & 3-layer HSTU + 9-layer AT & 0.1020 & 0.0795 & -0.33\% & -0.35\% \\
\bottomrule
\end{tabular}
\caption{Scaling comparison between vanilla HSTU and attention truncation at 3072 sequence length. Model is better when $\Delta$ NE is negative.}
\label{tab:hstu_vs_attntrunc_scaling_3k}
\end{table*}

\begin{algorithm*}[ht]
\caption{Load-Balanced Stochastic Length}
\label{alg:lbsl}
\begin{algorithmic}[1]
\Require World size $R$; warmup steps $T_{\text{warm}}$; recalibration interval $T_{\text{recal}}$;
load exponent $\gamma \in (1,2)$; SL parameter $\alpha$; SL sampling length $\ell_{\text{SL}}$; batch size $b$.
\State Initialize target load $\bar{\ell} \gets 0$ and $\ell_r \gets 0$ for all ranks $r\in \{1, \ldots, R\}$
\For{training step $t = 1,2,\dots$}
    \ForAll{ranks $r \in \{1,\dots,R\}$ \textbf{in parallel}}
        \State Receive local batch $\mathcal{B}_r$ with raw lengths $\{n_u\}_{u\in \mathcal{B}_r}$
        \State Compute $\ell_r \gets \ell_r + $\textsc{StandardSL\_Loads}$(\mathcal{B}_r;\alpha;\gamma)$ \Comment{pre-truncation proxy}
        \If{$t \le T_{\text{warm}}$}
            \State Apply \textsc{StandardSL}$(\mathcal{B}_r;\alpha)$ to truncate examples
        \Else
            \State $\mathcal{U}_r \gets \emptyset$; $s \gets 0$ \Comment{$\mathcal{U}_r$: untruncated set}
            \State Compute weights $p_u \gets \textsc{SLWeight}(n_u;\alpha)$ for all $u\in \mathcal{B}_r$
            \State Draw a weighted random permutation $\pi$ of $\mathcal{B}_r$ without replacement using $p_u$
            \For{$u$ in order $\pi$}
                \If{$s + (n_u^{\gamma} - \ell_{\text{SL}}^\gamma) \le \bar{\ell} - b \cdot \ell_{\text{SL}}^\gamma$}
                    \State $\mathcal{U}_r \gets \mathcal{U}_r \cup \{u\}$; $s \gets s + (n_u^{\gamma} - \ell_{\text{SL}}^\gamma)$
                \EndIf
            \EndFor
            \State Keep all $u \in \mathcal{U}_r$ \textbf{unsampled}
            \State For each $u \in \mathcal{B}_r \setminus \mathcal{U}_r$, apply the same sampling rule as \textsc{StandardSL}$(u;\alpha)$
        \EndIf
    \EndFor
    \If{$t = T_{\text{warm}}$}
        \State All-reduce to obtain mean load $\bar{\ell} \gets \frac{1}{RT_{\text{warm}}}\sum_{r=1}^{R}\ell_r$
        \State $\ell_r \gets 0$ for all ranks $r\in \{1, \ldots, R\}$
    \ElsIf{$(t \bmod T_{\text{recal}})=0$}
        \State All-reduce to obtain mean load $\bar{\ell} \gets \frac{1}{RT_{\text{recal}}}\sum_{r=1}^{R}\ell_r$
        \State $\ell_r \gets 0$ for all ranks $r\in \{1, \ldots, R\}$
    \EndIf
\EndFor
\end{algorithmic}
\end{algorithm*}

\section{Other System Optimizations}\label{other_system}

\begin{figure*}[t]
\begin{lstlisting}[caption={ULTRA-HSTU pseudocode with activation rematerialiation.},label={lst:hstu_full_layer}]
class HSTULayerFunction(torch.autograd.Function):
    @staticmethod
    def forward(ctx, x, norm_w, linear_w_1, gated_norm_w, linear_w_2):
        normed_x = norm_forward(x, norm_w)
        u, v, q, k = silu_forward(addmm(normed_x, linear_w_1))
        attn = attention_forward(q, k, v)
        y = gated_norm_forward(attn, u, gated_norm_w)
        out = addmm(y, linear_w_2, bias=x)
        ctx.save_for_backward(x, u, attn, norm_w, linear_w_1, gated_norm_w, linear_w_2)
        return out
    @staticmethod
    def backward(ctx, dout):
        x, u, attn, norm_w, linear_w_1, gated_norm_w, linear_w_2 = ctx.saved_tensors
        y = gated_norm_forward(attn, u, gated_norm_w) # rematerialize y
        normed_x = norm_forward(x, norm_w)
        u, v, q, k = silu_forward(addmm(normed_x, linear_w_1)) # rematerialize u, v, q, k
        dy = addmm(dout, linear_w_2.T)
        d_linear_w_2 = addmm(y.T, dout)
        dx = dout
        dattn, du, d_gated_norm_w = gated_norm_backward(dy, attn, u, gated_norm_w)
        dq, dk, dv = attention_backward(dattn, q, k, v)
        duqkv = concat(du, dq, dk, dv)
        uqkv = concat(u, q, k, v)
        d_addmm = silu_backward(duqkv, uqkv)
        d_normed_x = addmm(d_addmm, linear_w_1.T)
        d_linear_w_1 = addmm(normed_x.T, d_addmm)
        dx, d_norm_w = norm_backward(d_normed_x, x, norm_w)
        dx += dout
        return dx, d_norm_w, d_linear_w_1, d_gated_norm_w, d_linear_w_2
\end{lstlisting}
\end{figure*}

\paragraph{Mix-precision serving} In model serving, sparse embedding features can dominate host-to-device transfer time under long sequences. We therefore quantize embedding tensors to INT4 and keep them in quantized form across the embedding lookup and transfer path, reducing transfer volume and alleviating the communication bottleneck. In addition, we use groupwise INT4 that leverages group-specific scaling factors to significantly reducing quality loss compared to a single scale per row, while still delivering substantial throughput gains.

\section{Efficiency Benchmarks}\label{eff_bench}
In this section, we provide benchmarks to test the system optimizations presented in Section~\ref{sec:system_opt}.

\subsection{Mixed Precision Benchmarks}

\paragraph{FP8 precision efficiency} We evaluate the performance impact of FP8 on the GEMMs in both the pre-attention and post-attention blocks, and summarize the speedups in Table~\ref{tab:fp8-pre}. A key difference between the two blocks is the bias format in GEMM: the pre-attention GEMM uses a 1D bias, while the post-attention GEMM uses a 2D bias. Since FP8 GEMM with 1D bias is already supported by Torch, we directly use the Torch kernel for the pre-attention GEMM. In contrast, Torch FP8 GEMM does not natively support the 2D bias case needed by post-attention; therefore, we develop a customized Triton FP8 GEMM kernel with native 2D-bias fusion, and report its kernel-level efficiency in Table~\ref{tab:fp8}.

\begin{table}[htbp]
  \begin{center}
    \begin{small}
      \begin{sc}
        \begin{tabular}{lcc}
          \toprule
          Part    & \shortstack{Quant-Fused \\ FP8}  & \shortstack{Quant-Separate \\ FP8} \\
          \midrule
          Pre-attn  & 3.19X & 2.34X   \\
          Post-attn  & 1.99X  & 1.01X \\  
          \bottomrule
        \end{tabular}
      \end{sc}
    \end{small}
  \end{center}
  \centering\caption{Performance comparison between FP8 GEMM with standard quantization versus FP8 GEMM with quantization fused into preceding kernels, reporting speedup of each approach relative to BF16.}
  \label{tab:fp8-pre}
\end{table}

In Table~\ref{tab:fp8}, we benchmark the operation $\mathbf{D}=\mathbf{A}\mathbf{B}+\mathbf{C}$ where $\mathbf{A}\in\mathbb{R}^{m\times k}$, $\mathbf{B}\in\mathbb{R}^{k\times n}$, and $\mathbf{C}\in\mathbb{R}^{m\times n}$, using matrix dimensions $(m,k,n)$ that reflect our model workload. In particular, we emphasize large leading dimensions $m$ induced by jagged, variable-length sequences, which dominate the compute cost in practice. The results show that fusing the 2D bias directly into the FP8 GEMM (Bias-Fused FP8) provides up to 1.75$\times$ speedup compared to using Torch FP8 GEMM followed by a separate bias addition (Bias-Split FP8), motivating our Triton implementation for post-attention.

Finally, Table~\ref{tab:fp8-pre} reports the speedup of the complete pre-attention and post-attention parts when switching from BF16 to FP8. We observe strong end-to-end gains from two sources: (1) higher-performance FP8 GEMM kernels (including our 2D-bias Triton kernel for post-attention), and (2) additional savings from fusing quantization into the kernels that precede GEMM, which reduces extra memory traffic and kernel launch overhead compared to performing quantization as a separate step.

\paragraph{Int4 quantization efficiency}

The impact of Int4 sparse embedding quantization on model serving efficiency is summarized in Table~\ref{tab:int4}. Applying 4-bit quantization to sparse embeddings reduces host-to-device data-transfer latency by around 40\% 
and increases peak queries per second (QPS) by over 20\%.
In addition, we observed negligible differences in online model accuracy after applying 4-bit quantization.

\begin{table}[ht]
  \begin{center}
    \begin{small}
      \begin{sc}
        \begin{tabular}{lcc}
          \toprule
          Dtype    & Latency  & {Peak QPS} \\
          \midrule
          Int8  & 13ms & 3.6K   \\
          Int4  & 7.9ms ($\downarrow 40\%$)  & 4.4K ($\uparrow 22\%$) \\  
          \bottomrule
        \end{tabular}
      \end{sc}
    \end{small}
  \end{center}
  \centering\caption{Impact of sparse embedding datatypes over model serving efficiency performances. The latency of embedding lookup is measured at 3.5K QPS, and peak QPS refers to the maximum QPS at a end-to-end latency budget of 80ms. All results are collected from a single H100 host.}
  \label{tab:int4}
\end{table}

\begin{figure*}[htbp]
    \centering
    
    \begin{subfigure}[b]{0.30\textwidth}
        \centering
        \includegraphics[width=\textwidth]{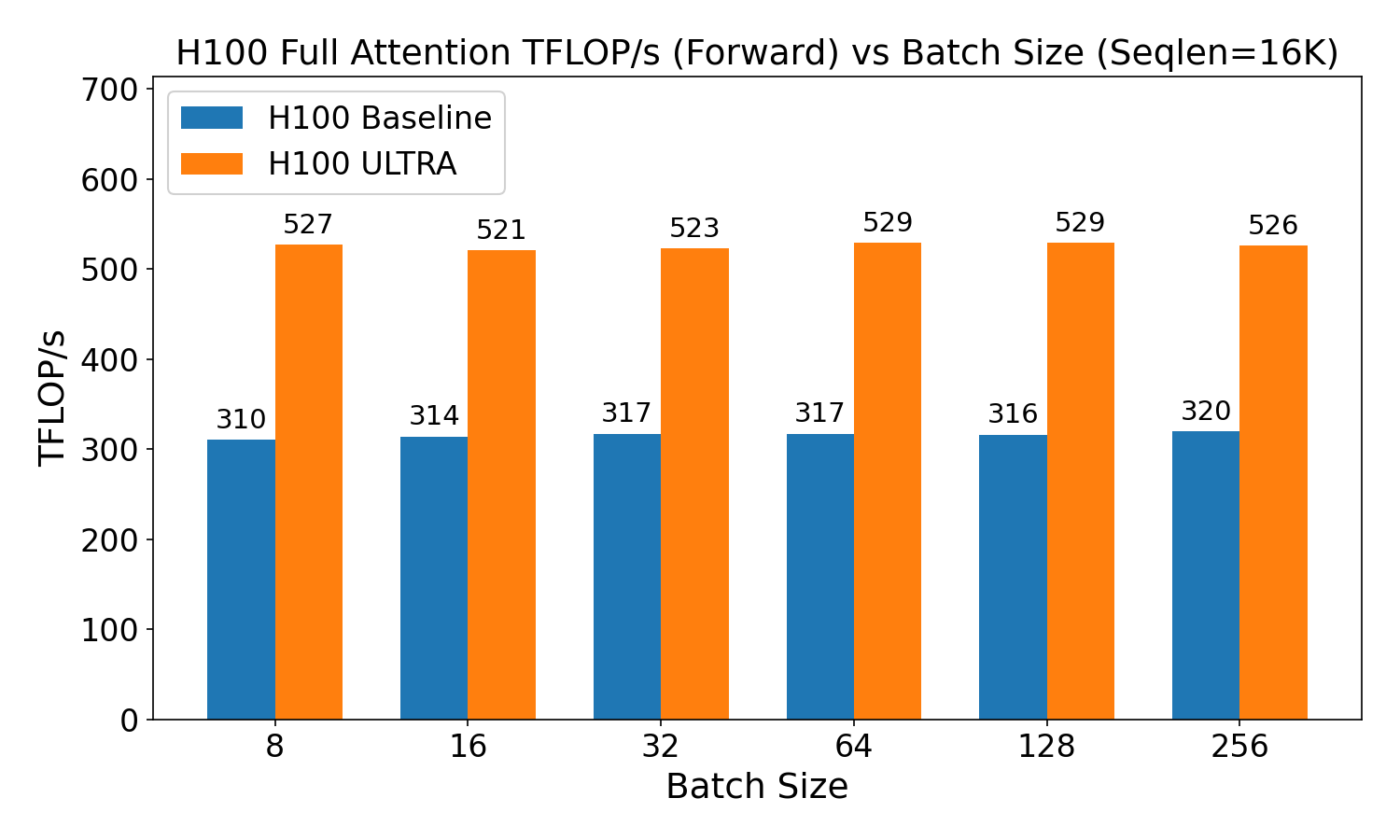}
        \caption{Forward, H100}
        \label{fig:attention_plot1}
    \end{subfigure}
    \begin{subfigure}[b]{0.30\textwidth}
        \centering
        \includegraphics[width=\textwidth]{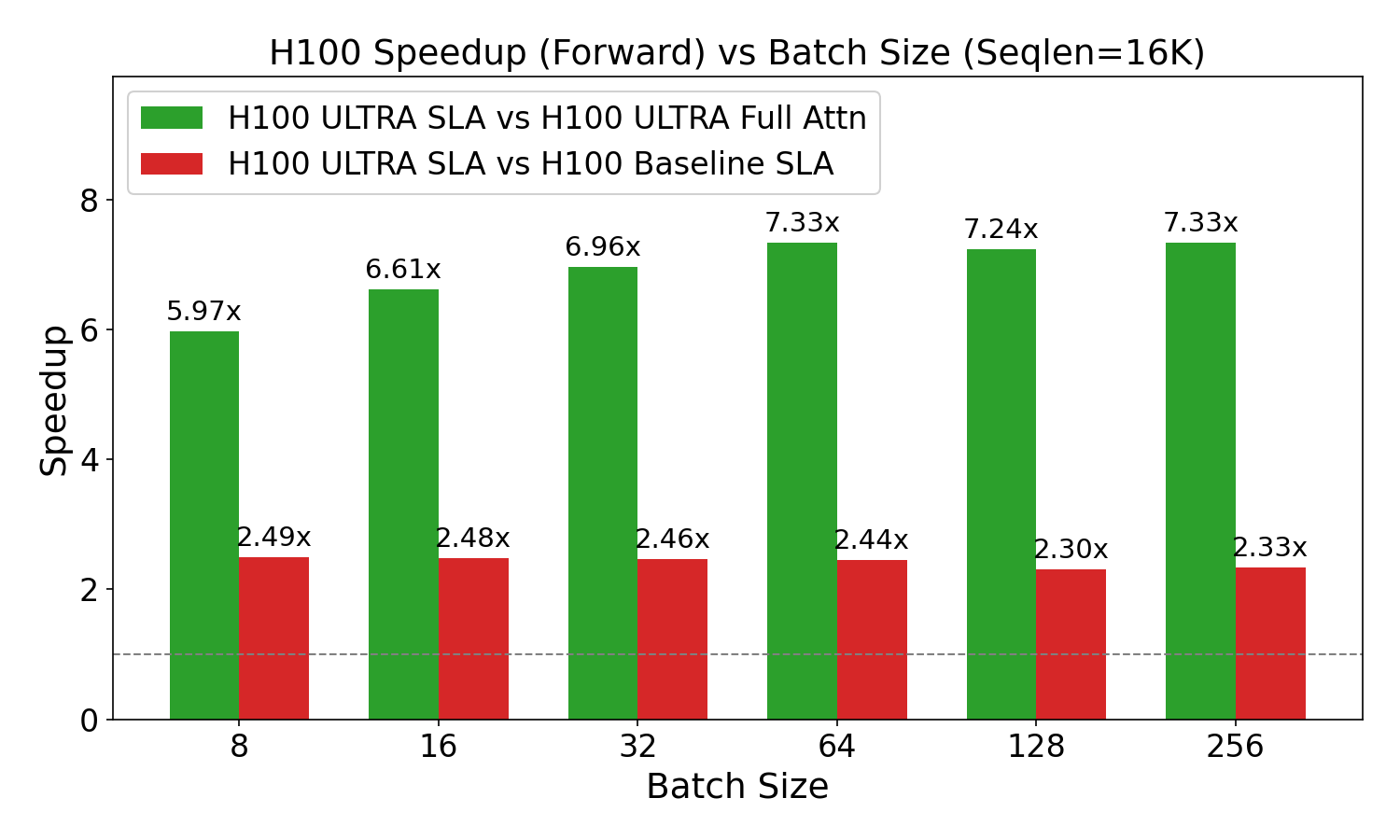}
        \caption{Forward, H100}
        \label{fig:attention_plot5}
    \end{subfigure}
    \begin{subfigure}[b]{0.36\textwidth}
        \centering
        \includegraphics[width=\textwidth]{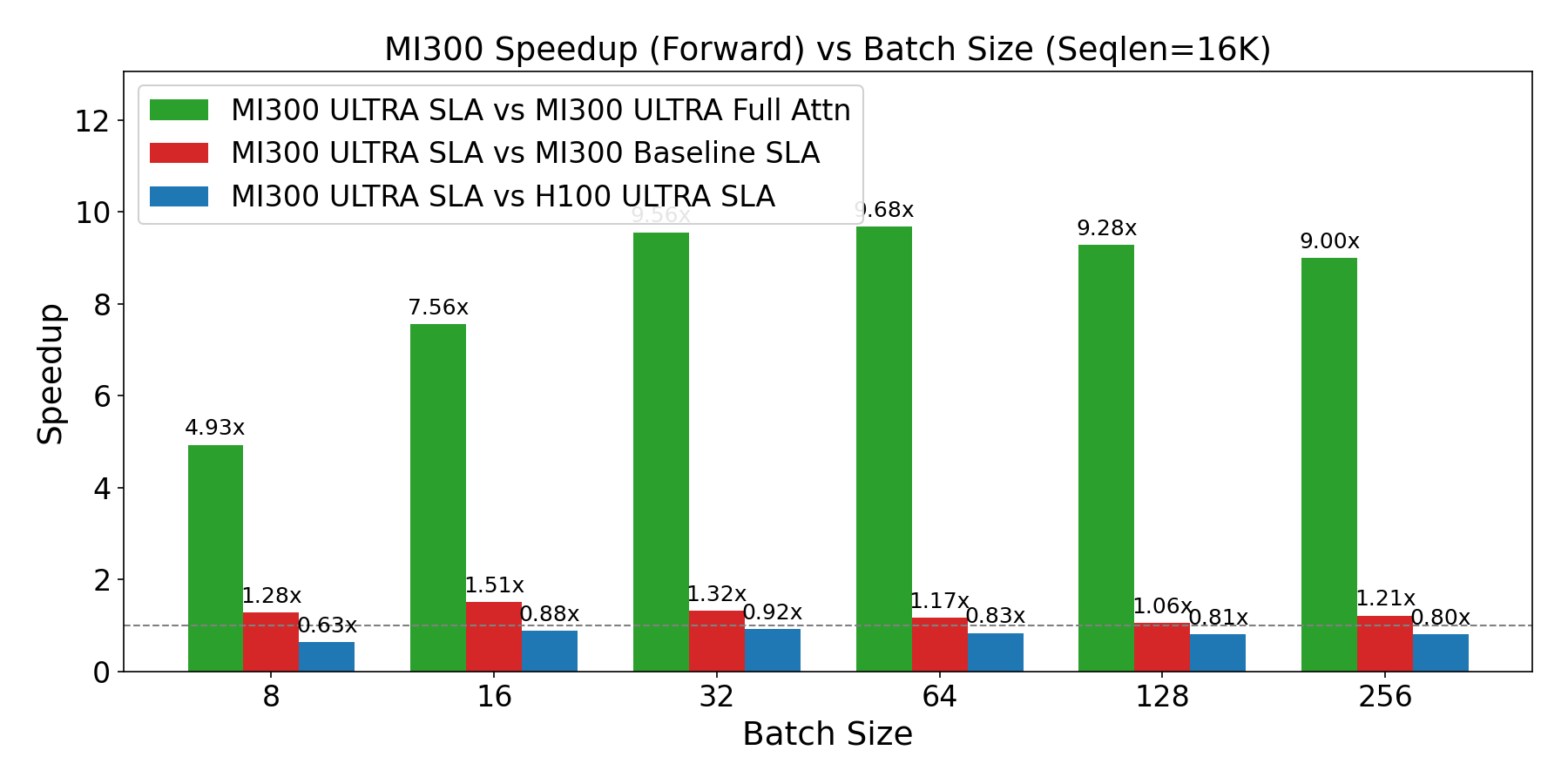}
        \caption{Forward, Mi300}
        \label{fig:attention_plot9}
    \end{subfigure}

    \begin{subfigure}[b]{0.30\textwidth}
        \centering
        \includegraphics[width=\textwidth]{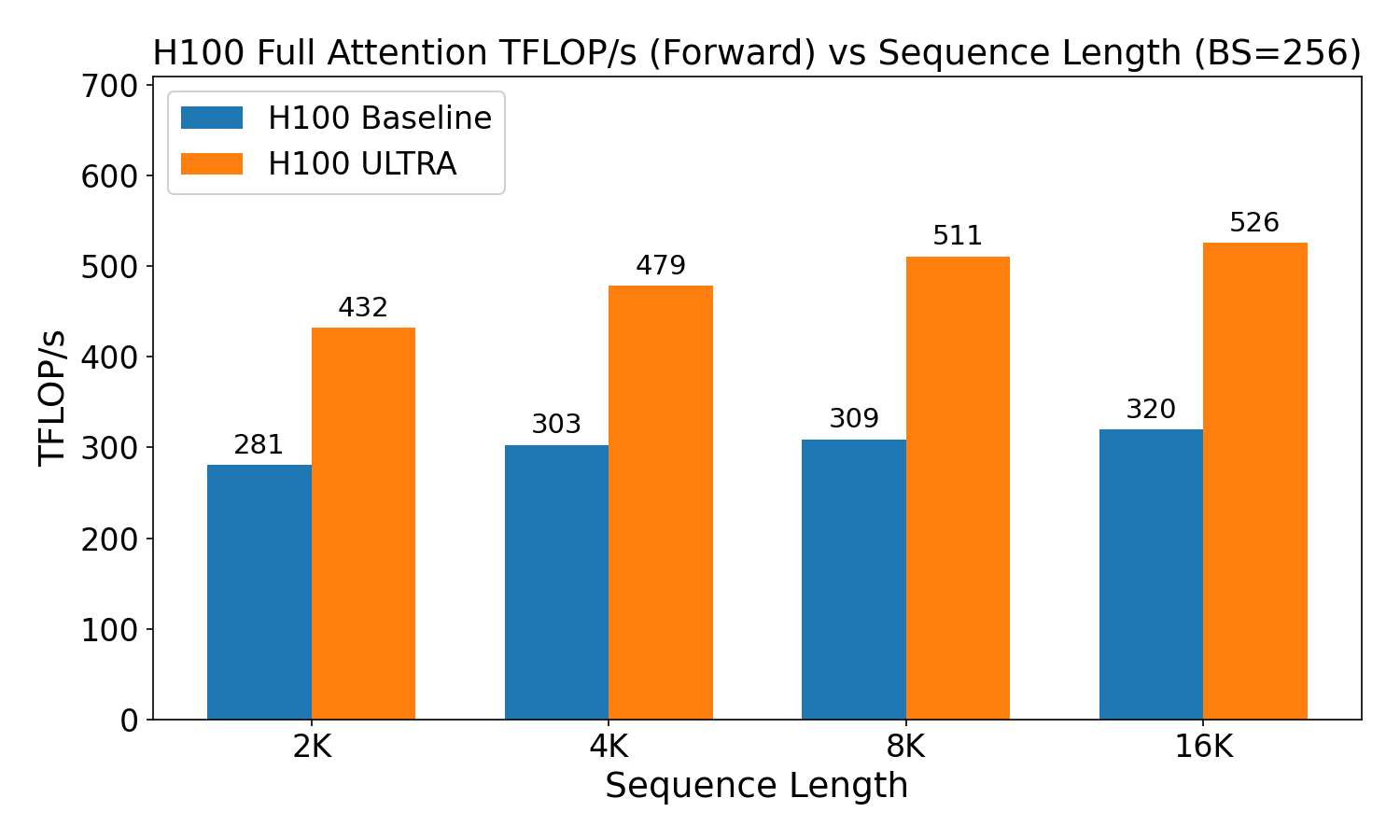}
        \caption{Forward, H100}
        \label{fig:attention_plot2}
    \end{subfigure}
    \begin{subfigure}[b]{0.30\textwidth}
        \centering
        \includegraphics[width=\textwidth]{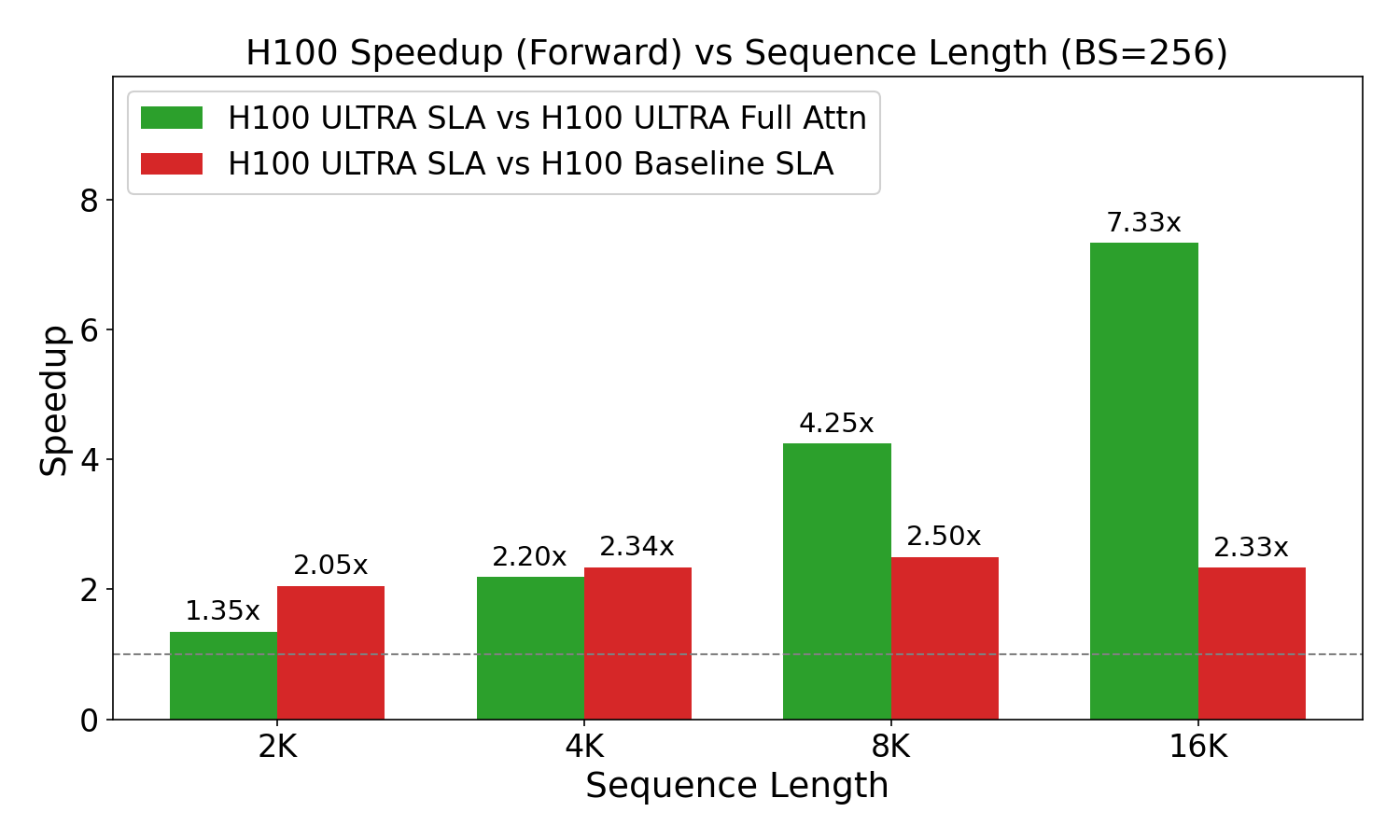}
        \caption{Forward, H100}
        \label{fig:attention_plot6}
    \end{subfigure}
    \begin{subfigure}[b]{0.36\textwidth}
        \centering
        \includegraphics[width=\textwidth]{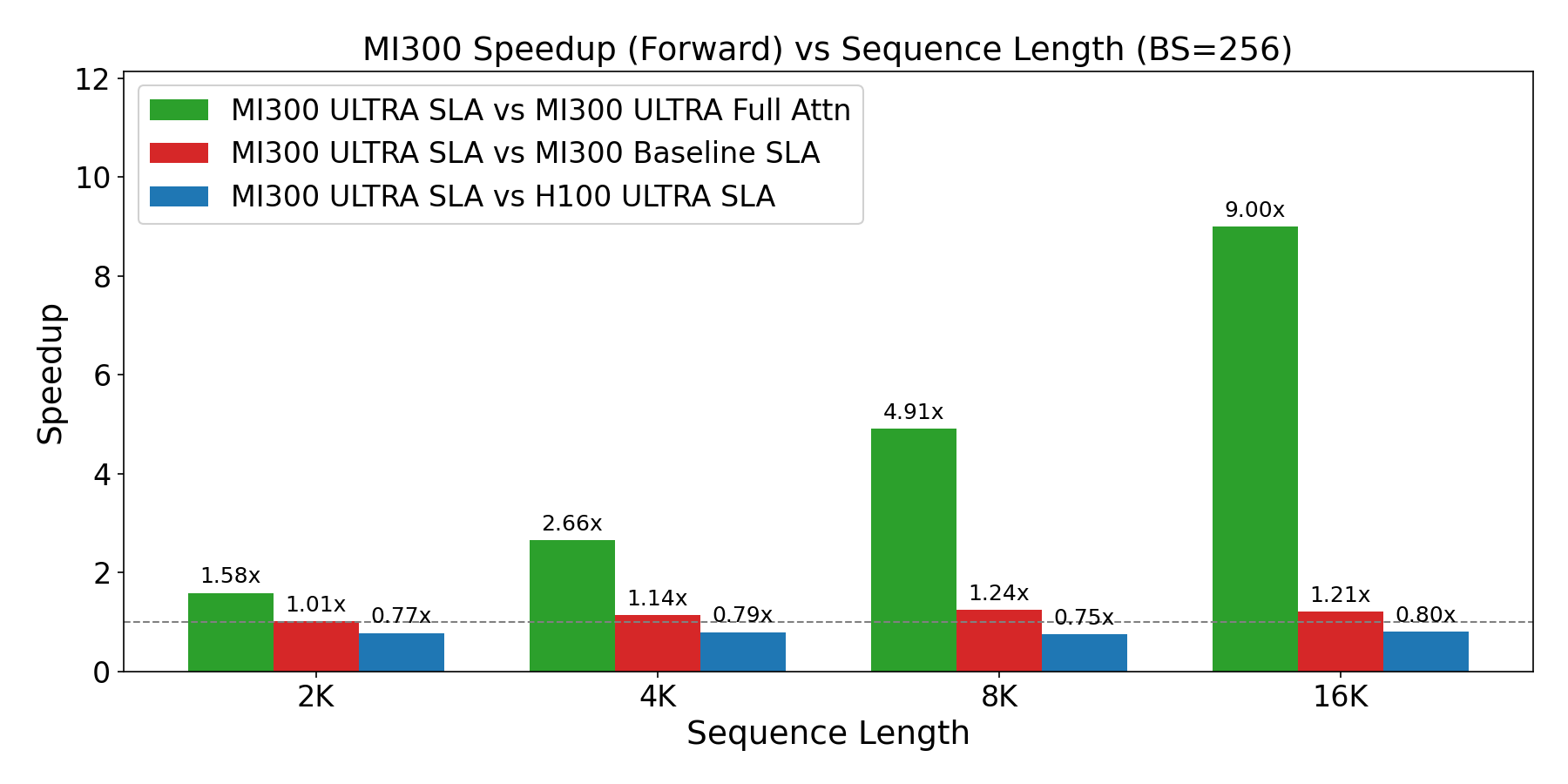}
        \caption{Forward, Mi300}
        \label{fig:attention_plot10}
    \end{subfigure}

    \begin{subfigure}[b]{0.24\textwidth}
        \centering
        \includegraphics[width=\textwidth]{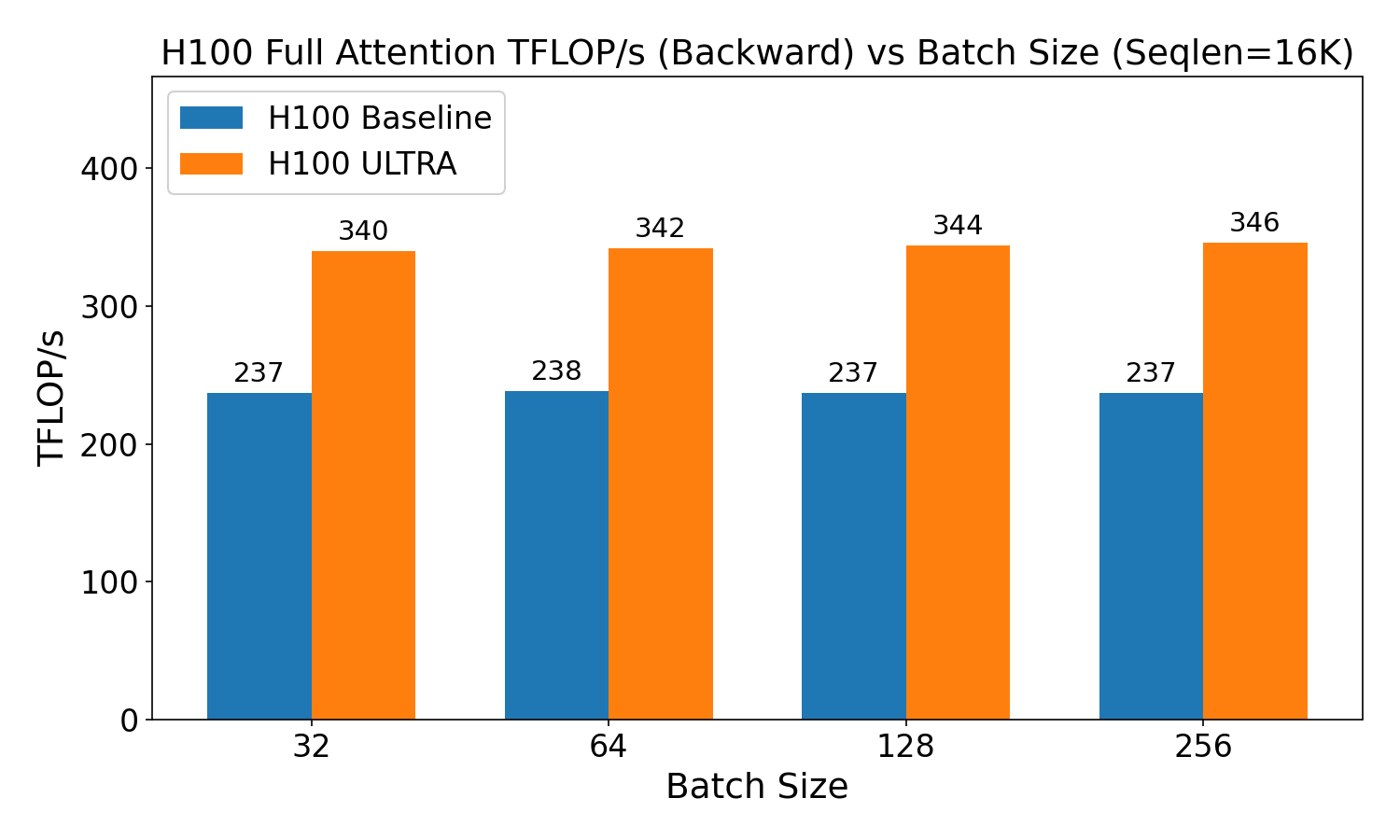}
        \caption{Backward, H100}
        \label{fig:attention_plot3}
    \end{subfigure}
    \begin{subfigure}[b]{0.24\textwidth}
        \centering
        \includegraphics[width=\textwidth]{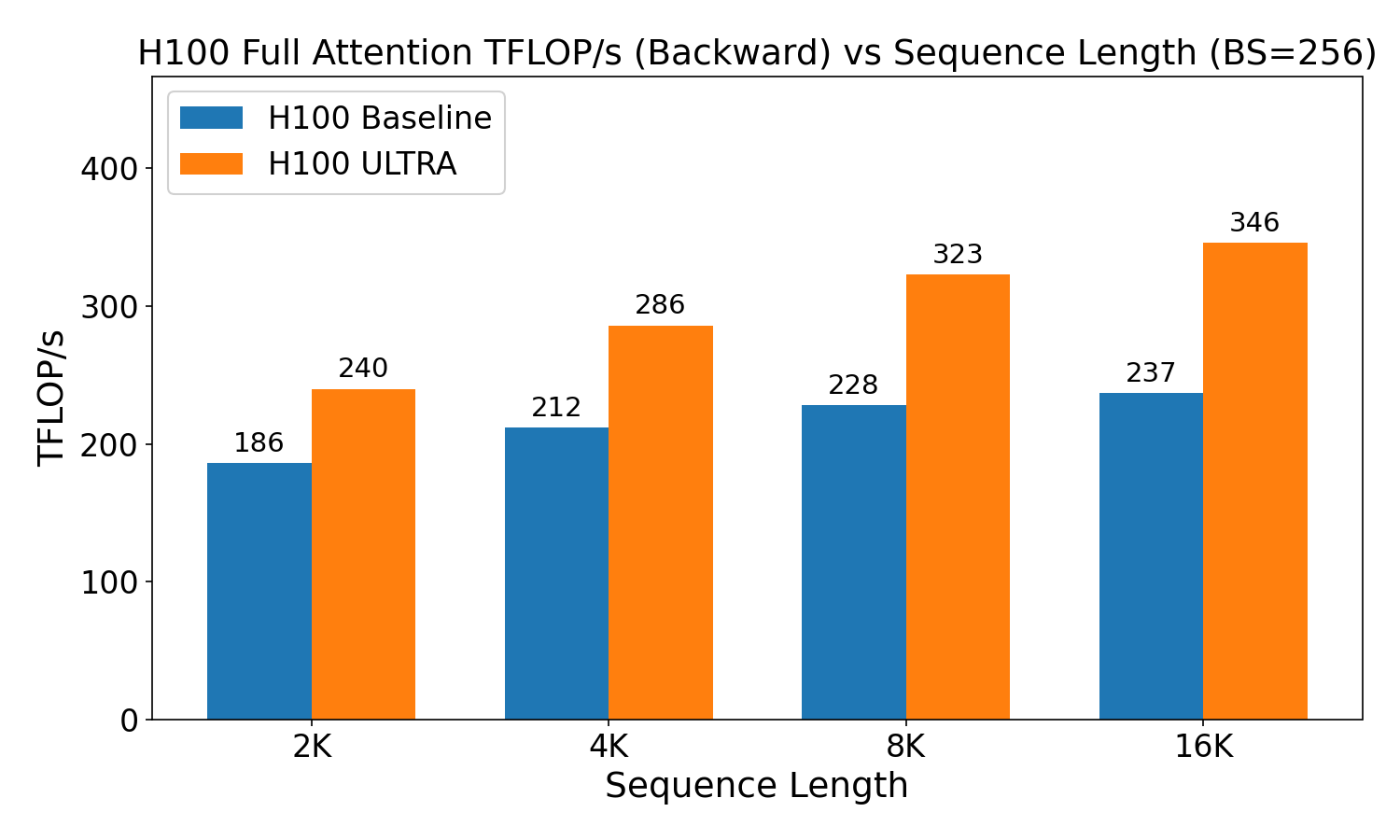}
        \caption{Backward, H100}
        \label{fig:attention_plot4}
    \end{subfigure}
    \begin{subfigure}[b]{0.24\textwidth}
        \centering
        \includegraphics[width=\textwidth]{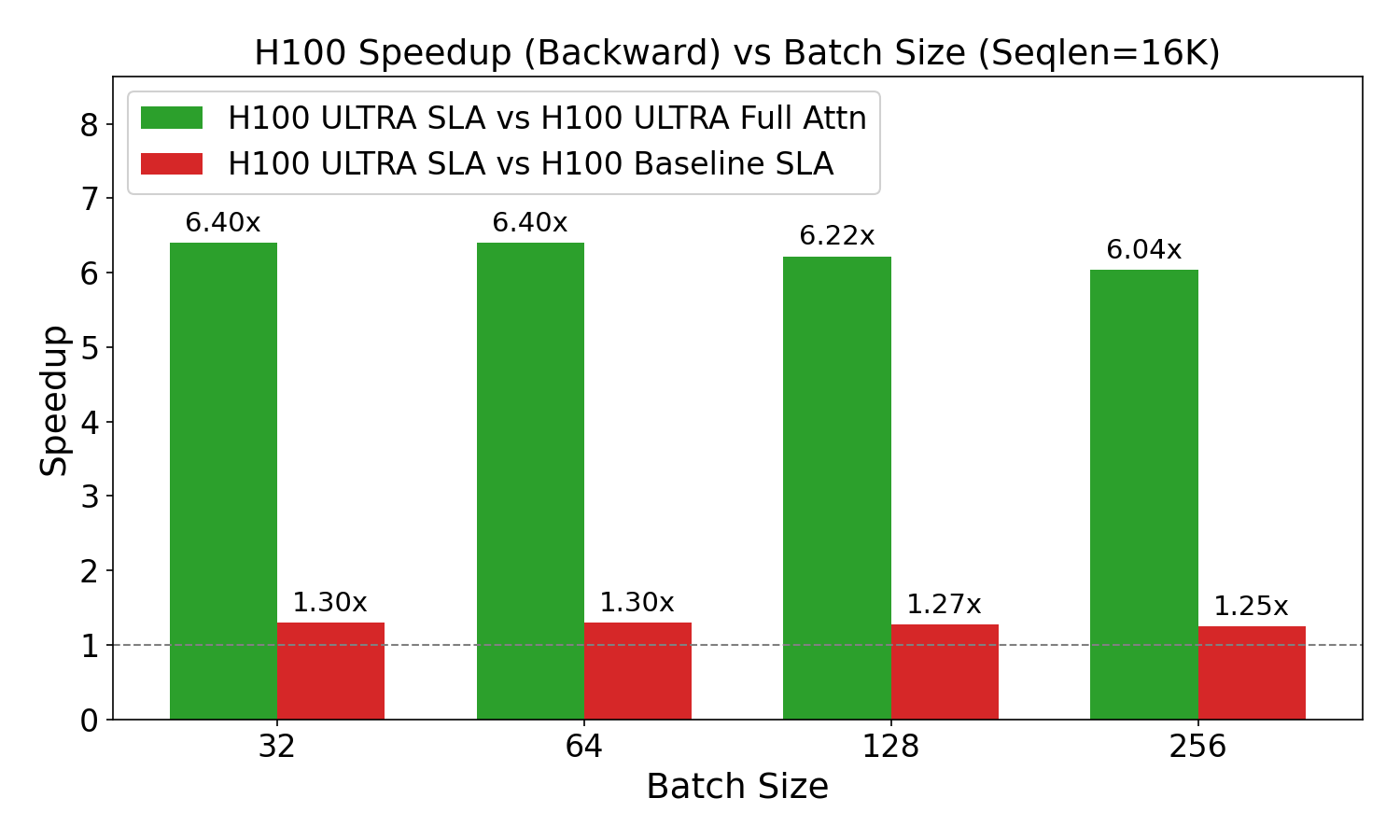}
        \caption{Backward, H100}
        \label{fig:attention_plot7}
    \end{subfigure}
    \begin{subfigure}[b]{0.24\textwidth}
        \centering
        \includegraphics[width=\textwidth]{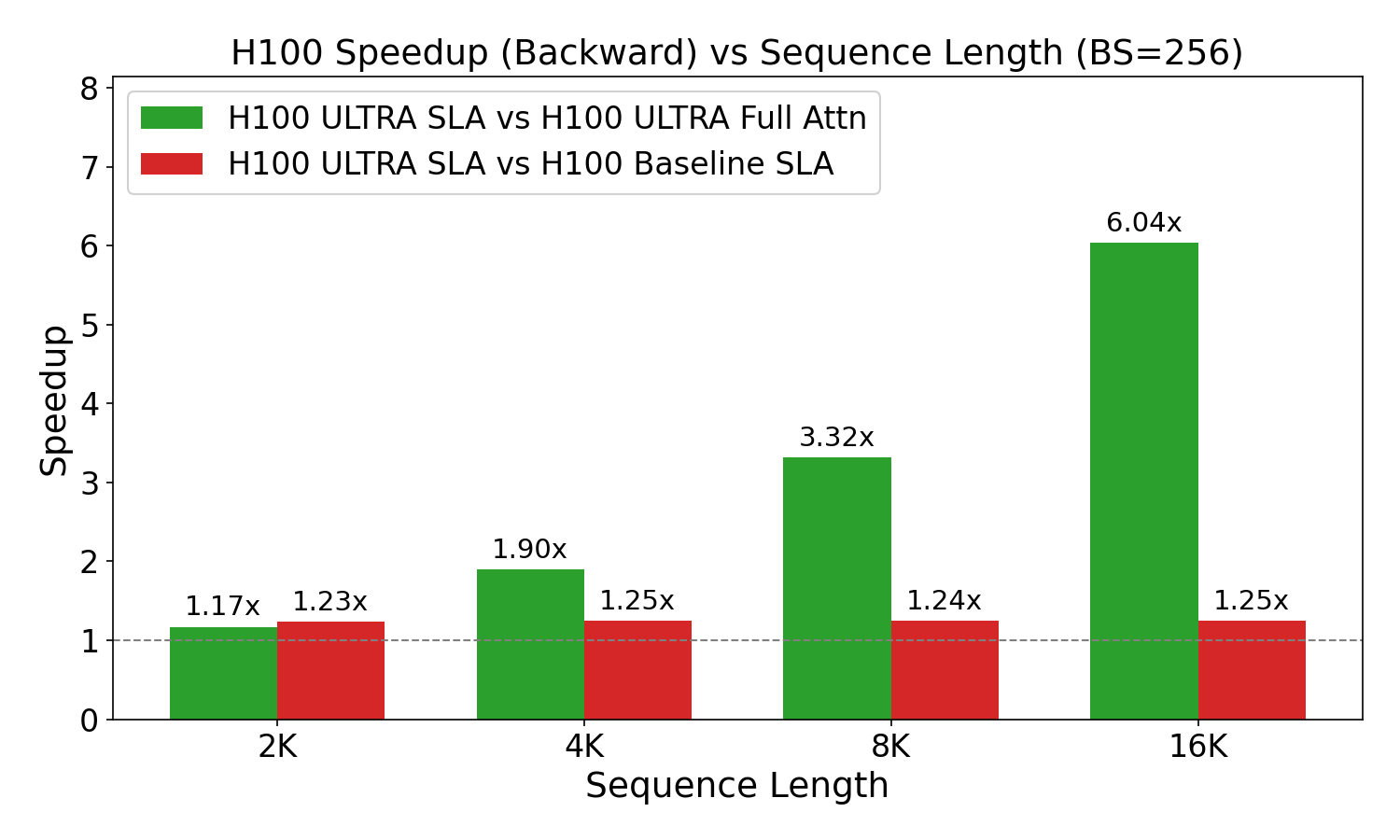}
        \caption{Backward, H100}
        \label{fig:attention_plot8}
    \end{subfigure}
    
    \caption{Performance comparison of ULTRA vs Baseline for both H100 and Mi300. ULTRA uses FlashAttention-V3-style algorithm, while baseline uses Triton implementation with FlashAttention-V2-style algorithm.
    (a)(d)(g)(h): full attention TFLOP/s for forward and backward attention kernels on H100. (b)(e)(i)(j): Speedup of ULTRA SLA over two baselines on H100: full attention with ULTRA implementation and SLA with baseline implementation. (c)(f): Speedup of ULTRA SLA over three baselines on Mi300.}
    \label{fig:all_attention_plots}
\end{figure*}

\subsection{Attention Kernel Benchmarks}

We present attention-kernel efficiency benchmarks in Figure~\ref{fig:all_attention_plots}, comparing our optimized implementation against the FlashAttention-V2 baseline on both NVIDIA H100 and AMD MI300 GPUs. We evaluate two settings: semi-local attention (SLA) and causal attention.

On H100, for causal attention, the ULTRA implementation sustains over 520 TFLOP/s at a 16K sequence length, delivering a $1.64\times$ speedup over the baseline. For SLA, across a range of batch sizes and sequence lengths, our kernels consistently achieve higher throughput and provide up to a 
$2.5\times $ speedup.

On MI300, Figure~\ref{fig:all_attention_plots} reports forward-pass kernel performance. Our ULTRA kernels deliver up to a $1.51\times$
 speedup over the FlashAttention-V2-based implementation. Relative to ULTRA on H100, ULTRA on MI300 achieves up to a $0.92\times$
 throughput ratio at 16K sequence length under small batch sizes. These results highlight our targeted efforts to enable efficient AMD inference for large-scale models.

\section{Scaling Laws for ULTRA-HSTU}\label{app:power-scaling-law-of-ultra-hstu}

\begin{figure*}[bt]
  \begin{center}
    \centerline{\hfill\includegraphics[width=0.8\columnwidth]{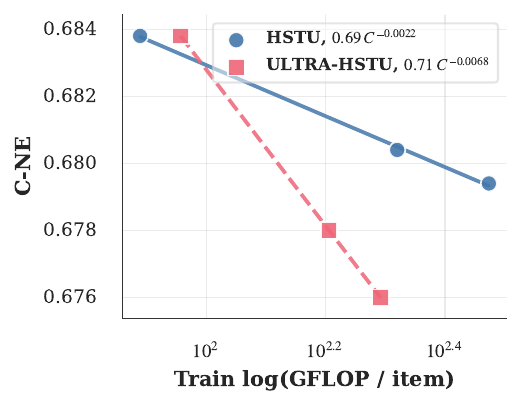}\hfill \includegraphics[width=0.8\columnwidth]{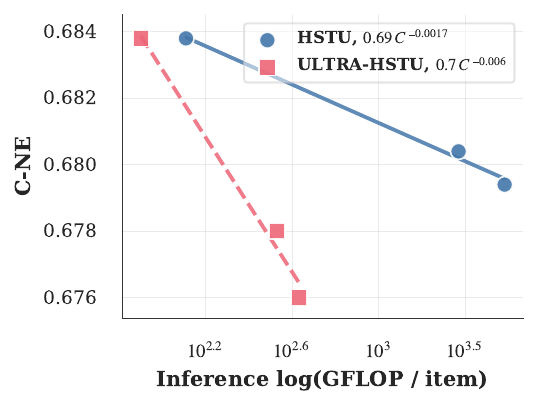}\hfill}
    \caption{Overall Compute Scaling Law: We compare the scaling performance of ULTRA-HSTU against vanilla HSTU with respect to training FLOP (left) and inference FLOP (right). 
    ULTRA-HSTU achieves $3.09\times$ and $3.52\times$ improvements in the training and inference compute scaling law exponents, respectively.} 
    \label{fig:power_full_scaling}
  \end{center}
\end{figure*}

\begin{figure*}[htbp]
  \begin{center}
    \centerline{
      \includegraphics[width=0.6\columnwidth]{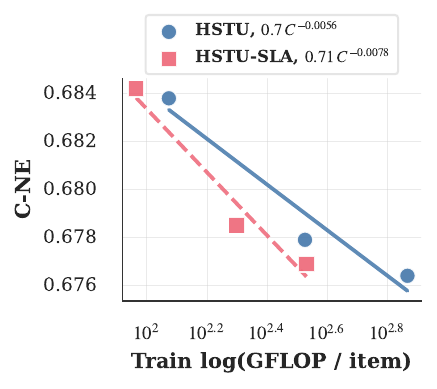}\hfill
      \includegraphics[width=0.6\columnwidth]{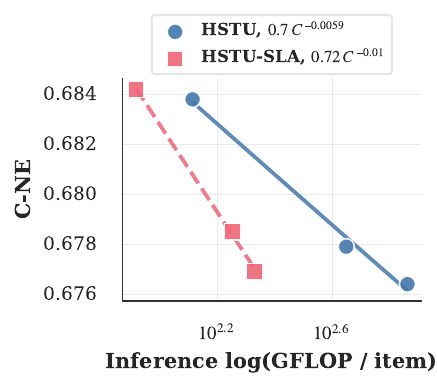}\hfill
      \includegraphics[width=0.6\columnwidth]{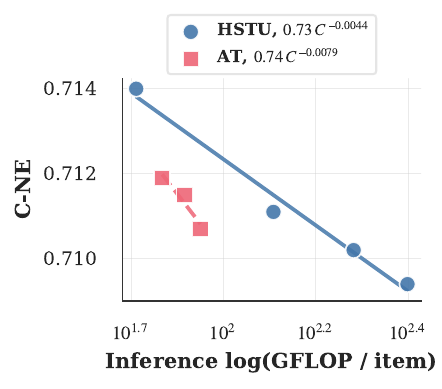}
    }
    \caption{Ablation study on scaling: Semi-Local Attention (SLA) improves the training scaling exponent by $1.39\times$ (left) and the inference scaling exponent by $1.69\times$ (middle). Attention Truncation Scaling Law: combining self-attention and attention truncation mechanisms yields a $1.8\times$ improvement in inference compute scaling (right).}
    \label{fig:sla_scaling}
  \end{center}
  \vspace{-1em}
\end{figure*}

In this section we analyze the compute scaling laws for the approaches
proposed in the present work. To do so we assume that the NE as a function 
of the compute follows a power-law of the form~\citep{kaplan2020scaling, hoffmann2022training},
\begin{equation}
    L(C) = \alpha C^{-\beta},
    \label{eq:scaling-power-law}
\end{equation}
where we have have assumed NE metrics $L(C)\to 0$ as computational budget $C \to \infty$. 
In general this will cause us to systematically underestimate the true scaling law 
exponent by the factor,
\begin{equation}
   \hat{\beta}  = \beta^*\left(1 - \frac{L_\infty}{L}\right) 
\end{equation} 
where $\hat{\beta}$ indicates our estimate for the true parameter
$\beta^*$ and $L_\infty$ is the irreducible error on the data.
The reason for this assumption is that it allows us 
to keep $\alpha$ and $\beta$ linear in  log-log space without having to estimate the irreducible error term. As we show below, this assumption also ensures that estimates for scaling improvement are conservative.

Consider the estimated scaling ratio between two models,
\begin{equation}
    \frac{\hat{\beta_1} }{\hat{\beta_2}} = 
    \frac{{\beta_1^*} }{{\beta_2^*}} \frac{\left(1 - L_\infty/L_1\right)}
    {\left(1-L_\infty/L_2\right)} = \frac{{\beta_1^*} }{{\beta_2^*}} R,
\end{equation}
where $R$ is a correction factor to the scaling 
ratio estimate. The scaling ratio tells us how much model 1 improves
on model 2's scaling curve.  
When $L_1 < L_2$ then $R < 1$.
When model 1 also achieves improved scaling (which is always
true for models with lower loss for the compute regions we 
consider) our estimate for the the scaling ratio is conservative.

\textbf{Interpreting scaling law exponent improvements.}
While improvements to scaling law exponents may appear modest at first glance, their impact compounds dramatically as compute budgets grow. To see this, consider two models with scaling laws $L_1(C) = \alpha C^{-\beta_1}$ and $L_2(C) = \alpha C^{-\beta_2}$, where $\beta_1 = k \beta_2$ for some improvement factor $k > 1$. To achieve the same loss with model 2 that model 1 achieves with compute $C$, we require,
\begin{equation}
    C_2 = C^k,
\end{equation}
meaning that the compute advantage grows as a polynomial with the scaling ratio. 
For example, a $2\times$ improvement in the scaling exponent implies that the baseline model requires {quadratically} more compute to match the improved model's performance. 

\textbf{Overall ULTRA-HSTU scaling performance.}
In Figure~\ref{fig:power_full_scaling}, we plot the fitted compute scaling laws for ULTRA-HSTU versus HSTU as a function of train and inference FLOP, respectively.
We see that ULTRA-HSTU improves the scaling exponent by $2.08\times$ compared to HSTU with respect 
to training FLOP and $4.59\times$ with respect to inference FLOP.

\textbf{Semi-Local attention scaling performance.}
We next isolate the contribution of Semi-Local Attention (SLA) to the overall scaling improvements. Figure~\ref{fig:sla_scaling} show the scaling performance of SLA with respect to training and inference FLOP. Our approach achieves an improvement to the scaling law exponent of $1.39\times$ and $1.69\times$ with respect to training and inference FLOP, respectively.

\textbf{Attention truncation scaling performance.}
Finally, we analyze the scaling behavior of our attention truncation approach. While improvements to the 
scaling law exponent with respect to training FLOP are minor, Figure~\ref{fig:sla_scaling} shows we achieve an 
improvement to the inference  FLOP exponent of $1.8\times$.

\end{document}